\documentclass[review]{elsarticle}

\usepackage{lineno,hyperref,epstopdf}

\journal{Computational Statistics \& Data Analysis}
\usepackage{amssymb, amsmath, booktabs, color}
\newtheorem{thm}{Theorem}
\newtheorem{lem}{Lemma}
\newtheorem{prop}{Proposition}

\newtheorem{rem}{Remark}

\newcommand{\be}{\begin{eqnarray}}
\newcommand{\ee}{\end{eqnarray}}
\newcommand{\bq}{\begin{eqnarray*}}
\newcommand{\eq}{\end{eqnarray*}}

\renewcommand{\baselinestretch}{1.4}

\newcommand{\bb}[1]{\boldsymbol{#1}}

\begin{document}

\begin{frontmatter}

\title{A general Monte Carlo method for multivariate goodness--of--fit testing applied to elliptical families\tnoteref{mytitlenote}}
\tnotetext[mytitlenote]{Supplementary material providing the R code and additional simulation results is provided online.}

%
\author[BNUZ]{Feifei Chen}

\author[UoS]{M. Dolores Jim\'enez--Gamero}

\author[NKUA,NWU] {Simos Meintanis\fnref{myfootnote}}
\fntext[myfootnote]{On sabbatical leave from the University of Athens.}

\author[BNUZ]{Lixing Zhu\corref{mycorrespondingauthor}}
\cortext[mycorrespondingauthor]{Corresponding author}
\ead{lzhu@hkbu.edu.hk}

\address[BNUZ]{Center for Statistics and Data Science, Beijing Normal University, Zhuhai, China}
\address[UoS]{Department of Statistics and Operations Research, University of Seville,  Spain}
\address[NKUA]{Department of Economics, National and Kapodistrian University of Athens, Athens, Greece}
\address[NWU]{Pure and Applied Analytics, North--West University, Potchefstroom, South Africa}

\begin{abstract}
A general and relatively simple method for construction of multivariate goodness--of--fit tests is introduced. The proposed test is applied  to elliptical distributions.
The method is based on a characterization of probability distributions via their characteristic function. The consistency and other limit properties of the new test statistics are studied.
Also in a simulation study the proposed tests are compared with earlier as well as more recent competitors.
\end{abstract}


\end{frontmatter}


\section{Introduction}
Let  $\bb{X}\in \mathbb R^p, \ p\geq 1$, be an arbitrary random vector with distribution function (DF) $F_{\bb{X}}(\bb{x})=\mathbb P(\bb{X}\leq  \bb{x})$. We consider the problem of goodness--of--fit testing with null hypothesis $F_{\bb{X}}=F_0$ where $F_0$ may be either a fixed distribution (simple hypothesis) or a family of distributions indexed by a parameter (composite hypothesis). While goodness--of--fit has a long history dating back to Pearson's chi--squared test (see the volume edited by \citealp{HBNM} for a review of goodness--of--fit including historical accounts), the approach followed herein is relatively recent. Specifically we employ the characteristic function (CF) of ${\bb X}$ under the null hypothesis and measure its distance from the empirical CF computed from the sample in order to decide whether our data  are compatible with the null hypothesis. In this connection, there is a line of research with CF--based testing  methods that are developed by adapting the original idea of Pearson's chi--squared test to the context of CFs; see for instance \cite{KK81}, \cite{ES86}, \cite{Fan1997} and \cite{MK99}. However the by far most favoured approach is that of continuous-type procedures, starting with the contribution  of \cite{Epps1983} for testing univariate normality, that was followed by a series of papers by N. Henze and coauthors addressing the problem of multivariate normality; see \cite{Baringhaus1988}, \cite{HZ90}, and \cite{Henze1997}. The reader is also referred to the review articles of \cite{Henze2002} and \cite{Ebner2020} that give an up to date picture of the wide range of procedures available for this particular problem of testing for normality. Outside the multivariate Gaussian context, the range of goodness--of--fit procedures available is certainly much more limited; see   \cite{Fragiadakis2011}, \cite{MH10} and \cite{MNT} for testing for the Laplace, the skew--normal and the stable, distribution, respectively; see also \cite{JG2009} for a general approach.

Compared to the aforementioned procedures the approach followed herein is different in that it frames any such test as a two--sample test, whereby the test statistic measures distance between  the empirical CF of the data at hand and an empirical CF computed from  a Monte Carlo sample generated under the null hypothesis. In doing so, and borrowing from earlier CF--based methodology we are able to express this test statistic in a convenient analytical form that facilitates computations, its asymptotics are derived under the null  hypothesis as well as under alternatives, while its actual implementation is carried out by using well established resampling procedures.

The remainder of this work unfolds as follows. In Section \ref{sec2} we provide a characterization that provides the background for our tests. In Section \ref{sec3} we introduce our method for goodness--of--fit testing as applied to simple null hypotheses under test, without estimated parameters. Section \ref{sec4} extends the new tests to the case of a composite null hypothesis with unknown parameters within the family of elliptical distributions. Asymptotic properties of the tests are provided in Section \ref{sec5}. Section \ref{sec6} specifies the actual implementation of the new test procedure.  In Section \ref{sec7} a simulation study is presented whereby the suggested method is applied on several popular null hypotheses under test, including comparisons with alternative methods. A real--data example is also included. The paper concludes in Section \ref{sec8} with a discussion and outlook. Technical proofs are deferred to the Appendix in Section \ref{sec9}. An online Supplement contains the \textsf{R} code and some extra simulation results.


\section{Background} \label{sec2}
In this section we prove, and discuss related aspects regarding, a characterization that motivates our test statistic. To this end, consider the characteristic function (CF) $\varphi_{\bb{X}}(\bb{t})=\mathbb E(e^{{\rm{i}} \bb{t}^\top  \bb{X}})$, $\bb{t} \in \mathbb R^p$, of $\bb X$, where  ${\rm{i}}=\sqrt{-1}$ and the superscript $\top$ means transposition of column vectors and matrices. Consider the population Fourier--type discrepancy measure
\begin{equation} \label{tw}
T_{w}= \int |\varphi_{\bb{X}}(\bb{t})-\varphi_0(\bb{t})|^2w(\bb{t}) {\rm{d}}\bb{t}
\end{equation}
between the CF of $\bb{X}$ and the CF $\varphi_0(\cdot)$ of an independent random vector $\bb{X}_0$ with fixed DF  $F_0(\cdot)$, where an unspecified integral denotes integration over $\mathbb{R}^{p}$.  In \eqref{tw}, $w(\cdot)$ denotes a non-negative weight function which will be further specified below and $| \cdot |$ stands for the modulus of a complex number.

Recall that for $z_1,z_2 \in \mathbb C$, the set of complex numbers, with Cartesian coordinates $z_m= {\rm{Re}}(z_m)+{\rm{i}} \: {\rm{Im}}(z_m), \ m=1,2$, we have $|z_1-z_2|^2=|z_1|^2+|z_2|^2-2 ({\rm{Re}}(z_1){\rm{Re}}(z_2)+{\rm{Im}}(z_1){\rm{Im}}(z_2))$. Thus for fixed $\bb{t}$, and since  $|\varphi_{\bb X}(\bb t)|^2=\varphi_{\bb X}(t)\varphi_{-\bb Y}(\bb t)=\varphi_{\bb X-\bb Y}(\bb t)$ (and likewise for $|\varphi_{0}(\bb t)|^2$) the CF point discrepancy in  \eqref{tw} may be written as
\begin{eqnarray} \label{dt}
\nonumber
|\varphi_{\bb{X}}(\bb{t})-\varphi_0(\bb{t})|^2&=& \varphi_{\bb X-\bb Y}(\bb t)+\varphi_{\bb X_0-\bb Y_0}(\bb t)- 2 {\rm{Re}}(\varphi_{\bb X-\bb X_0} (\bb t)) \\ \label{phi2} &=& \mathbb E\left( \cos(\bb{t}^\top (\bb{X}-\bb{Y}))+\cos(\bb{t}^\top (\bb{X}_0-\bb{Y}_0))-2\cos(\bb{t}^\top (\bb{X}-\bb{X}_0))\right),
\end{eqnarray}
where $\bb{Y}$ and $\bb{Y}_0$ is a pair of independent vectors which represent independent copies of $\bb{X}$ and $\bb{X}_0$, respectively, and the equality in \eqref{phi2} following by the fact that $\bb X-\bb Y$ and $\bb X_0-\bb Y_0$ are symmetric around zero.
From  \eqref{dt} and using Fubini's theorem we have
\begin{equation} \label{tw2}
T_w= \mathbb{E} \left( \int \big( \cos({\bb{t}}^\top (\bb{X}-\bb{Y}))+
\cos({\bb{t}}^\top (\bb{X}_0-\bb{Y}_0))-2\cos({\bb{t}}^\top (\bb{X} - \bb{X}_0)) \big) w(\bb{t}){\rm{d}}\bb{t} \right).
\end{equation}

Now suppose that $w(\bb{t})$ is the density of a spherical distribution in $\mathbb  R^p$. Then it is well known that the CF corresponding to $w(\cdot)$ can be written as $\Psi(\|\bb{x}\|^2)$
where $\Psi(\cdot)$ is called the ``kernel" of the specific family of spherical distributions and $\| \cdot \|$ stands for the Euclidean norm; see \cite{Fang1990}.

Thus if we apply $\Psi(\cdot)$ on \eqref{tw2} we have $T_w={\cal{E}}_\Psi(F,F_0)$ with
\begin{equation}  \label{tw1}
{\cal{E}}_\Psi(F_{\bb{X}},F_0)= \mathbb E\left( \Psi(\|\bb{X}-\bb{Y}\|^2)+\Psi(\|\bb{X}_0-\bb{Y}_0\|^2)-2\Psi(\|\bb{X}-\bb{X}_0\|^2)\right),
\end{equation}
and by invoking the uniqueness property of CFs we have the following characterization:

\begin{prop}\label{prop1} For any given kernel $\Psi$ and fixed DF $F_0$, the quantity defined by \eqref{tw1} satisfies  ${\cal{E}}_\Psi(F_{\bb{X}},F_0)\geq 0$ for every DF $F_{\bb{X}}(\cdot)$, while  ${\cal{E}}_\Psi(F_{\bb{X}},F_0)=0$ if and only if $F_{\bb{X}}\equiv F_0$.
\end{prop}

In this paper we propose goodness--of--fit tests that make use of Proposition \ref{prop1}. Understandably we favor  simple kernels $\Psi(\cdot)$ that lead to  test statistics that are straightforward to compute. In this connection, the most well known subfamily of multivariate distributions in $\mathbb R^p$ that possess simple kernels is the family of spherical and elliptical distributions. Here we will consider kernels from this subfamily, and since as already mentioned all considered distributions under test will also be specific members of the spherical/elliptical subfamily in what follows we briefly define this subfamily.

As already implicit, a random variable $\bb X\in \mathbb R^p$ is said to follow a spherical distribution if its CF satisfies $\varphi_{\bb X}(\bb t)=\Psi(\|\bb t\|^2)$ for some function $\Psi(\xi)$ of a scalar variable  $\xi$. Popular members of the spherical subfamily are the standard normal distribution corresponding to $\Psi(\xi)=e^{-\xi/2}$, the stable family of distributions (see \citealp{Nolan2013}) with $\Psi(\xi)=e^{- \xi^{b/2}}, \ b \in (0,2]$, and the generalized Laplace distribution (see \citealp{Kozubowski2013}) with $\Psi(\xi)=(1+ \xi)^{-b}, \ b>0$. The multivariate Student--t as well as the family of Kotz--type distributions are also members of the spherical subfamily of multivariate distributions in the standard Euclidean space, but these are more convenient to define via their densities rather than their CFs; see \cite{Kotz2004} and \cite{Nadarajah2003}, respectively. Each  spherical distribution gives rise to a corresponding  elliptical distribution by means of the transformation $\bb X \mapsto \bb \delta +\bb V^{1/2} \bb X$, with CF given by $e^{{\rm{i}}\bb t^\top \bb \delta}\Psi(\bb t^\top \bb V \bb t)$, where $\bb \delta \in \mathbb R^p$ is a location parameter and the $(p\times p)$ positive definite matrix  $\bb V$ represents scatter.  Further properties of spherical/elliptical distributions that are necessary for our procedure will be provided below. For more information  we refer \cite{Kelker70} and \cite{Fang1990}.



\begin{rem}
The quantity figuring in the right--hand side of equation \eqref{tw1} is a special case of the squared maximum mean discrepancy defined in Lemma 6 of \cite{GBRSS}.  Thus Proposition \ref{prop1} refers to Theorem 5 in \cite{GBRSS} (proved in the general setting of reproducing kernel Hilbert spaces), in effect putting emphasis on the fact that  CFs of spherical distributions lead to, in the terminology of \cite{GBRSS}, universal kernels, i.e. kernels that safeguard test consistency. In fact \cite{MXZ} already mention the normal and (generalized) Laplace as universal kernels. The underlying idea can be traced back to \cite{BF04} and \cite{SR05}, that state certain related characterizations, but these are stated in terms of quantities other than the  CF. Later \cite{SR13} revisited these characterizations and defined the so--called energy distance statistics thereof, which relate to equation \eqref{tw} in that energy statistics also use CFs. The difference is that the latter statistics, even in their generalized versions,  use a very specific non--integrable weight function, rather than a density,  and thus impose certain moment conditions on the underlying random variable that restricts application to light--tailed distributions. In this connection, the reader is referred to \cite{SSGF} for an equivalence result connecting energy--based statistics and maximum mean discrepancy statistics.
\end{rem}

\section{Preamble: Goodness--of--fit tests for simple hypothesis} \label{sec3}
In this section we consider the simple hypothesis of testing goodness--of--fit for a fixed, but otherwise arbitrary, distribution in $\mathbb R^p$ without estimated parameters, starting with the standard Gaussian law which is by far the most popular special case. To this end, let ${\cal{X}}_n:=(\bb{X}_1,\ldots,\bb{X}_n)$  be independent and identically distributed (i.i.d.) copies of $\bb{X}$, and suppose that under the null hypothesis
\be \label{null}
{\cal{H}}_{0}: \varphi_{\bb{X}}(\bb{t})=\varphi_0(\bb{t}), \quad \forall \bb{t} \in \mathbb R^p,
\ee
the law of $\bb{X}$ is fully specified.
For instance, in the case of the standard Gaussian law we have  $\varphi_0(\bb{t})=e^{-\|\bb{t}\|^2/2}$. Moreover certain simplifications occur if we use the standard Gaussian density as weight function. An unbiased estimator of  $n{\cal{E}}_\Psi(F_{\bb{X}},F_0)$, defined in  \eqref{tw1}, can be built as follows: replace the first expectation in \eqref{tw1} by the corresponding U--statistic,  the second expectation by its analytical expression, while for the third expectation, we first calculate the expectation with respect to $\bb{X}_0$, and then replace that expectation, with respect to $\bb{X}$,  by the corresponding U--statistic. This gives rise to the test statistic
\begin{equation*}  \label{tsgauss}
T^{(G)}_{n}= \frac{2}{n-1} \sum_{j<k} e^{-\|\bb{X}_j-\bb{X}_k\|^2/2}+\frac{n}{3^{p/2}}-2^{1-(p/2)}  \sum_{j=1}^n  e^{-\|\bb{X}_j\|^2/4},
\end{equation*}
which is in all essential aspects equivalent  to the Baringhaus--Henze--Epps--Pulley (BHEP)  test for the Gaussian distribution in the case of specified parameters; see \cite{Ebner2020} for a review of the BHEP test.

This approach of setting the weight function equal to the corresponding density when a spherical distribution is under test, as appealing as it may  be, it is nevertheless ad hoc and only convenient for a selective class of testing problems. Otherwise it is generally difficult to pinpoint a specific weight function for which the expectations occurring  in \eqref{tw1} can be explicitly computed for each testing problem. A partial solution is proposed in  \cite{Meintanis2014}, which is based on certain data transformation. Nevertheless, such transformation cannot be easily applied in many cases. For this reason,  here we advocate an entirely Monte Carlo--based automatic approach that can be used for any testing problem at hand, be it for a univariate or multivariate distribution, discrete or continuous, and not only for the simple hypothesis ${\cal{H}}_0$ but also for the composite hypothesis with estimated parameters. The only requirement is that we should be able to draw Monte Carlo samples from the distribution under test.

Specifically given an arbitrary testing problem such as that figuring in \eqref{null}, i.e. given any goodness--of--fit problem with a fixed distribution under test, we suggest to implement a test via a given kernel $\Psi(\cdot)$, and by estimating $n\mathcal{E}_\Psi(F_{\bb{X}},F_0)$ 
by
\begin{equation}  \label{ts}
T^{(\Psi)}_{n}= \frac{2}{n-1} \sum_{j<k} \Big[ \Psi(\|\bb{X}_j-\bb{X}_k\|^2)+\Psi(\|\bb{X}_{0,j}-\bb{X}_{0,k}\|^2)\Big] - \frac{2}{n-1} \sum_{j,k=1}^n \Psi(\|\bb{X}_j-\bb{X}_{0,k}\|^2),
\end{equation}
where ${\cal{X}}_{0,n}:=(\bb{X}_{0,1},\ldots,\bb{X}_{0,n})$ denote independent copies of $\bb{X}_0$, i.e. ${\cal{X}}_{0,n}$ are artificial data generated under the null hypothesis ${\cal{H}}_0$.
From Proposition  \ref{prop1}, a reasonable test should reject the null hypothesis   ${\cal{H}}_{0}$ in \eqref{null}  for large values of $T^{(\Psi)}_{n}$.
 Note that if the kernel $\Psi(\cdot)$ is simple enough, then the test statistic figuring in \eqref{ts} may be computed in a straightforward manner.

We close this section by noting that Proposition \ref{prop1} and the corresponding statistic figuring in \eqref{ts} are quite general imposing minimal restrictions on the family being tested. All that is needed is that we should be able to draw Monte Carlo samples from the distribution under test and, in the case of composite hypotheses that follows in the next section, possess a reasonable estimation method for the unknown distributional parameter. Thus the proposed method applies to arbitrary continuous distributions in $\mathbb R^p$ (symmetric or not symmetric) as well as discrete distributions, and could in principle even be extended to more general manifolds. Nevertheless we will herein restrict our attention to tests for certain popular cases within the family of spherical/elliptical distributions, a multivariate family which is clearly one of the the most important in both theory and applications.

\section{Goodness--of--fit tests for composite hypothesis} \label{sec4}
We now turn to the problem of composite goodness--of--fit testing. In doing so we focus on the case of elliptical distributions under test and the null hypothesis
 \be \label{null2}
 {\cal{H}}_{\bb{\vartheta}}: \varphi(\bb{t})=\varphi_{\bb{\vartheta}}(\bb{t}), \quad \forall \bb{t}\in \mathbb R^p, \ \mbox{for some} \ \bb{\vartheta} \in \Theta,
 \ee
where $\varphi_{\bb{\vartheta}}(\cdot)$ denotes the CF of a specific parametric family of elliptical distributions, say  ${\cal{F}}_{\bb{\vartheta}}$, admitting a parameterization in terms of the parameter $\bb{\vartheta}:=(\bb{\bb{\delta}}, \bb{V})$. In what follows $\varphi_0(\bb{t})=\Psi_0(\|\bb{t}\|^2)$ stands for the CF under the null hypothesis ${\cal{H}}_{\bb{\vartheta}}$ in \eqref{null2} corresponding to  $\bb{\vartheta}=\bb{\vartheta}_0:=(\bb{0}, {\rm{I}}_p)$, where ${\rm I}_p$ denotes the identity matrix of order $p$ and all $\bb{0}$s appearing in the manuscript are vectors or matrices  with the appropriate dimension and all elements equal to $0$. We note that the family
${\cal {F}}_{\bb{\vartheta}}=\{ F_{\bb{\vartheta}}, \, \bb{\vartheta} \in \Theta,\}$ may involve extra parameters which here are considered as fixed (known).

In this connection it is well-known  that if $F_{\bb{X}} = F_{\bb{\vartheta}}$, for some $\bb{\vartheta}=(\bb{\bb{\delta}}, \bb{V})$, then $F_{\bb{A X}+\bb{b}} =F_{\bb{\vartheta_{b,A}}}$,  with $\bb{\vartheta_{b,A}}=(\bb{A \bb{\delta}}+\bb{b}, \bb{A  V  A}^\top)$, for any  $(p\times p)$ non-singular matrix $\bb{A}$ and $\bb{b} \in \mathbb R^p$. This implies that the family of elliptical distributions ${\cal{F}}_{\bb{\vartheta}}$ is closed under affine transformations of the type $\bb{X}\mapsto  \bb{A  X}+ \bb{b}$. Consequently it is good statistical practice to require that any test statistic, say $T_{n}({\cal{X}}_n)$, also be invariant in the sense that $T_{n}({\cal{X}}_n)=T_{n}(\bb{A} {\cal{X}}_n+\bb{b})$, for each  $(\bb{b},\bb{A})$; see \cite{Henze2002} for a detailed discussion of the notion of affine invariant tests in the context of testing for normality. One compromise to affine invariance is to impose the condition that $T_{n}({\cal{X}}_n)$, although not affine invariant, it has a distribution that is asymptotically affine invariant, i.e. the limit distribution of the test statistic is independent of $(\bb{\bb{\delta}},\bb{V})$. We will refer to such tests as asymptotically invariant--in--law tests.

With these considerations in mind, we suggest to test the null hypothesis $\cal{H}_{\bb{\vartheta}}$ figuring in \eqref{null2} by means of the test statistic
\begin{equation}  \label{ts1}
\widehat T^{(\Psi)}_{n}= \frac{2}{n-1} \sum_{j<k} \Big[\Psi(\|\widehat {\bb{X}}_j-\widehat {\bb{X}}_k\|^2)+\Psi(\|\bb{X}_{0,j}-\bb{X}_{0,k}\|^2) \Big] -\frac{2}{n-1} \sum_{j,k=1}^n\Psi(\|\widehat { \bb{X}}_j- \bb{X}_{0,k}\|^2),
\end{equation}
on the basis of standardized observations
\be \label{stand}
\widehat { \bb{X}}_j=\widehat {\bb{V}}^{-1/2}_n ( \bb{X}_j-\widehat {\bb{\bb{\delta}}}_n), \,\,  j=1,\ldots,n,
\ee
where ${\widehat {\bb{\vartheta}}}_n:=(\widehat {\bb{\bb{\delta}}}_n, \widehat {\bb{V}}_n)$ denotes an estimator of the parameter $\bb{\vartheta}=(\bb{\bb{\delta}},\bb{V})$ computed from  the data vector ${\cal{X}}_n$, ${\cal{X}}_{0,n}:=( \bb{X}_{0,1}, \ldots,  \bb{X}_{0,n})$ denotes a random sample from $ \bb{X}_0$, that has distribution function $F_{\bb{\vartheta}_0}$,  and 
$\widehat {\bb{V}}^{-1/2}_n$ stands for the unique symmetric square root of  $\widehat {\bb{V}}^{-1}_n$, which is tacitly assumed to exist. Under ${\cal{H}}_{\bb{\vartheta}}$ in
\eqref{null2},  the behaviour of $\widehat { \bb{X}}_1, \ldots, \widehat { \bb{X}}_n$ should imitate that of $ \bb{X}_{0,1}, \ldots,  \bb{X}_{0,n}$, at least for large samples. Thus, in view of Proposition \ref{prop1}, it is sensible to reject the null hypothesis  ${\cal{H}}_{\bb{\vartheta}}$  for large values of
$\widehat T^{(\Psi)}_{n}$. Next section gives a sound justification for this critical region and Section \ref{sec6} gives a procedure to approximate critical points.

We now discuss the estimation of $\bb{\vartheta}$ always keeping in mind the aforementioned invariance property of a test statistic. In this connection we assume that the estimator  $ {\widehat {\bb{\vartheta}}}_n:= {\widehat {\bb{\vartheta}}}_{n}({\cal{X}}_n)$ of  $\bb{\vartheta}$ satisfies,
\begin{equation}  \label{equiv}
\widehat {\bb{\bb{\delta}}}_{n}(\bb{A}{\cal{X}}_n+\bb{b})=\bb{A}\widehat {\bb{\bb{\delta}}}_{n}({\cal{X}}_n)+\bb{b},
\end{equation}
and
\begin{equation}  \label{inv}
\widehat{\bb{V}}_{n}(\bb{A}{\cal{X}}_n+\bb{b})=\bb{A}\widehat{\bb{V}}_{n}({\cal{X}}_n)\bb{A}^\top.
\end{equation}
Maximum likelihood estimators (MLEs)  satisfy \eqref{equiv}--\eqref{inv}. However the MLEs are often hard to compute and ensuring their uniqueness is also an issue; see \cite{Bilodeau1999}. On the other hand moment estimators (MEs) are more straightforward and typically also satisfy   \eqref{equiv}--\eqref{inv}. Specifically and assuming $\mathbb E\|\bb{X}\|^2<\infty$, we make use of the fact that if $\bb{X}$ has distribution function $F_{\bb{\vartheta}}$,  with $\bb{\vartheta}=(\bb{\bb{\delta}},\bb{V})$, then $\mathbb E(\bb{X})=\bb{\bb{\delta}}$ and $\mathbb V(\bb{X})=-2 \Psi'_0(0)\bb{V}$, where $\Psi_0(\cdot)$ stands for the kernel under the null hypothesis and  the symbol $\mathbb{V}$ denotes the covariance matrix of random vectors.
(Recall that if $\bb{X}$ follows a spherical distribution then $\varphi_X(\bb{t})=\Psi(\|\bb{t}\|^2)$, for some kernel $\Psi(\cdot)$). Thus the MEs may be obtained from the equations
\[
{\widehat {\bb{\delta}}}_n=\overline {\bb{X}}_n, \ \
{\widehat {\bb{V}}}_n=-\frac{\bb{S}_n}{2 \Psi'_0(0)},  \]
where  $\overline {\bb{X}}_n=n^{-1} \sum_{j=1}^n\bb{ X}_j$ and $\bb{S}_n=n^{-1}\sum_{j=1}^n (\bb{X}_j-\overline {\bb{X}}_n)(\bb{X}_j-\overline {\bb{X}}_n)^\top$ are the sample mean and the sample covariance matrix, respectively, and  $\Psi'_0(\xi)={\rm{d}} \Psi_0(\xi)/{\rm{d}}\xi$.

Our last assumption concerning the estimator  ${\widehat {\bb{\vartheta}}}_n=(\widehat {\bb{\delta}}_n,\widehat{\bb{V}}_n)$ is the so-called asymptotic Bahadur representation around the true value $ \bb{\vartheta}=({\bb{\delta}},\bb{V})$,
\begin{eqnarray}
\sqrt{n}\left(\widehat{{\bb{\delta}}}_n-{\bb{\delta}} \right) & = & n^{-1/2}  \sum_{j=1}^n   {\bb{\ell}}_{\bb{\vartheta}}( \bb{X}_j)+{\rm{o}}_{\mathbb P}(1), \,\, \mathbb E\{\bb{\ell}_{\bb{\vartheta}}( \bb{X}_1)\}=0, \,\,  \mathbb V\{\|\bb{\ell}_{\bb{\vartheta}}( \bb{X}_1)\|\}<\infty,\label{bahadur1} \\
\sqrt{n}\left(\widehat{\bb{V}}_n- \bb{V} \right) & = & n^{-1/2}  \sum_{j=1}^n  {\bb L}_{\bb{\vartheta}}( \bb{X}_j)+{\rm{o}}_{\mathbb P}(1),
\,\, \mathbb E\{{\bb L}_{\bb{\vartheta}}( \bb{X}_1)\}=0, \,\,  \mathbb V\{\|{\bb L}_{\bb{\vartheta}}( \bb{X}_1)\|\}<\infty. \label{bahadur2}
\end{eqnarray}
If $\bb{\vartheta}=\bb{\vartheta}_0$, then we simply write $\bb{\ell}_0$ and ${\bb L}_0$ for $\bb{\ell}_{\bb{\vartheta}}$ and ${\bb L}_{\bb{\vartheta}}$, respectively.

When we jointly assume that \eqref{equiv} and \eqref{bahadur1} (resp. \eqref{inv} and \eqref{bahadur2}) hold, it is tacitly assumed that
the function $\bb{\ell}_{\bb{\vartheta}}(\cdot)$ (resp. ${\bb L}_{\bb{\vartheta}}(\cdot)$) fulfills $\bb{\ell}_{\bb{\vartheta}}(\bb{AX}+\bb{b})=\bb{A}\bb{\ell}_{\bb{\vartheta}}(\bb{X})+\bb{b}$ (resp. ${\bb L}_{\bb{\vartheta}}(\bb{AX}+\bb{b})=\bb{A} {\bb L}_{\bb{\vartheta}}(X)\bb{A}^\top$).

In the next section we derive some asymptotic properties of the test statistic $\widehat T^{(\Psi)}_{n}$ under the null hypothesis as well as under alternatives. In this connection we show that the limit null distribution of the test statistic is independent of the true value of  $\bb{\vartheta}$, which entails that  $\widehat T^{(\Psi)}_{n}$ is asymptotically invariant--in--law.

\section{Asymptotics} \label{sec5}

This section studies some asymptotic properties of the test statistic   $\widehat T^{(\Psi)}_{n}$ defined  in \eqref{ts1}. Specifically, Subsection \ref{part.one} studies the limit behaviour of $\widehat T^{(\Psi)}_{n}/n$ showing that it converges to a non-negative quantity, which is equal to 0 if and only if the null hypothesis is true, thus giving a sound justification for the critical region intuitively given in the previous section.  Subsection
\ref{part.two} derives the asymptotic null distribution of the test statistic and shows that, for certain common choices of  $\widehat{{\bb{\delta}}}_n$ and $\widehat{\bb{V}}_n$, the asymptotic null distribution does not depend on ${\bb \vartheta}$. This result justifies the procedure proposed in Section \ref{sec6} to approximate the critical points. From the results in Subsections \ref{part.one} and \ref{part.two}, it follows that the proposed test is consistent against any fixed alternative.  Finally,  Subsection \ref{part.three} derives the asymptotic distribution of the test statistic under alternatives, which is useful for approximating the power and to  compare the behaviour of the proposal with the ``ideal" (but unfeasible, from a practical point of view) test whose test statistic calculates the expectations (with respect to ${\bb X}_0$) in  \eqref{tw1}. Before deriving those results, we first introduce some useful notation.

Notice that $\widehat T^{(\Psi)}_{n}$ is invariant under translation of the sample ${\cal{X}}_n$, i.e. it holds $\widehat T^{(\Psi)}_{n}({\cal{X}}_n+\bb{b})=\widehat T^{(\Psi)}_{n}({\cal{X}}_n)$, for each  $\bb{b}\in \mathbb R^p$,
and therefore the distribution of $\widehat T^{(\Psi)}_{n}$ does not depends on $\bb{\bb{\delta}}$.  Observe also that $\widehat T^{(\Psi)}_{n}(\bb{U}{\cal{X}}_n,\bb{U} {\cal{X}}_{0,n})=\widehat T^{(\Psi)}_{n}({\cal{X}}_n, {\cal{X}}_{0,n})$, for each $ \bb{U} \in \mathcal{O}_p$, where $\mathcal{O}_p$ is the set of $(p \times p)$ orthogonal matrices. As a consequence,
there is no loss of generality in assuming that ${\bb{\delta}}={\bb 0}$ and that $\bb{V}$ is a diagonal matrix.

For asymptotics it is convenient, by application of Lemma 1 in \cite{Alba2008}, to express the test statistic as
\begin{equation} \label{equality}
\widehat T^{(\Psi)}_{n}=\frac{n}{n-1}\left( \widehat T^{(\Psi)}_{2,n}-2\right),
\end{equation}
where
\[
\widehat T^{(\Psi)}_{2,n}=n\int |\varphi_n(\bb{t})-\varphi_{0,n}(\bb{t})|^2w(t){\rm{d}}\bb{t},
\]
with $w(\bb{t})$ being the density of a continuous spherically symmetric distribution with CF $\Psi(\|\bb{x}\|^2)$, and
\be \label{ecf}
\varphi_n(\bb{t})=\frac{1}{n}\sum_{j=1}^n e^{{\rm i}\bb{t}^\top \widehat {\bb{X}}_j}, \quad
\varphi_{0,n}(\bb{t})=\frac{1}{n}\sum_{j=1}^n e^{{\rm i}\bb{t}^\top  \bb{X}_{0,j}}
\ee
being the empirical CFs corresponding to  $(\widehat {\bb{X}}_1, \ldots, \widehat {\bb{X}}_n)$ and
$(\bb{X}_{0,1},\ldots, \bb{X}_{0,n})$, respectively.
%

Our study of asymptotics, will make use of equation \eqref{equality}. To this end, let $L_w^2$ stand for the space of all $L^2$ functions defined on the measure space $(\mathbb{R}^p, \mathcal{B}_p,\nu)$ taking values in $\mathbb{C}$, the set of complex numbers, where $\mathcal{B}_p$ denotes the $\sigma$-field of Borel subsets of
$\mathbb{R}^p$,  and the measure $\nu$ has density $w$:
${\rm{d}} \nu(\bb{u})=w(\bb{u}){\rm{d}}\bb{u}$, that is,
$L_w^2=\{ f:\mathbb{R}^p \mapsto \mathbb{C}: \|f\|^2_w=\int |f( \bb{u})|^2w(\bb{u}) {\rm{d}} \bb{u}<\infty\}$. Let $\langle \cdot , \cdot \rangle_w$ denote the scalar product in  the (separable) Hilbert space $L_{w}^2$. With this notation and taking into account that
\begin{equation}\label{sym}
 w(\bb{t})=w(-\bb{t}), \quad \forall \bb{t} \in \mathbb{R}^p,
\end{equation}
we have
\begin{equation}\label{T2}
\widehat T^{(\Psi)}_{2,n}=n\|G_n\|_w^2,
\end{equation}
where
\begin{equation}\label{Gn}
G_n(\bb{t})={\rm Re}(\varphi_n(\bb{t})) + {\rm Im}(\varphi_n(\bb{t})) - {\rm Re}(\varphi_{0,n}(\bb{t})) - {\rm Im}(\varphi_{0,n}(\bb{t})),
\end{equation}
Recall that $z={\rm Re}(z)+ {\rm i} \: {\rm Im}(z)$ is the Cartesian representation of  $z \in \mathbb{C}$.

\subsection{Stochastic limit of the test statistic} \label{part.one}

The following theorem  studies the limit  of  $\widehat T^{(\Psi)}_{n}/n$ without assuming any parametric form for the law of ${\bb X}$. In Theorem \ref{limit} as well as in subsequent results, a convergence in probability assertion  may be replaced by a corresponding almost sure assertion, provided that in the associated assumptions, convergence in probability is replaced by almost sure convergence. The following notation will be used:
${\rm{diag}}(a_1, \ldots, a_p)$ denotes a diagonal matrix of order $p$;
 $\overset{\mathbb{P}}{\to}$ denotes convergence in probability;  all limits in this paper are taken when  $n \rightarrow \infty$, $n$ denoting the sample size.

\begin{thm} \label{limit}
Let $\bb{X}_1, \ldots, \bb{X}_n$ be i.i.d. copies of the random variable $\bb{X} \in \mathbb{R}^p$, and assume that the estimator $\widehat {\bb{\vartheta}}_n=(\widehat{\bb{\delta}}_n,\widehat{\bb{V}}_n)$ satisfies
\begin{eqnarray*}
\widehat{\bb{\delta}}_n  &  \overset{\mathbb{P}}{\to} & \bb{\delta}=\bb{0}, \\
\widehat{\bb{V}}_n   & \overset{\mathbb{P}}{\to}& \bb{V}={\rm{diag}}(\lambda_1, \ldots, \lambda_p),
\end{eqnarray*}
for some $0<\lambda_j<\infty, \ j=1,\ldots,p$. Then provided that $\int \|\bb{t}\|^2w(\bb{t})d\bb{t}<\infty$,
$$\frac{\widehat T^{(\Psi)}_{n}}{n} \ \overset{\mathbb{P}}{\to} \ \|\varphi_{\bb{Y}}-\varphi_{0}\|_w^2:=\kappa,$$ 
where $\bb{Y}=\bb{V}^{-1/2}\bb{X}$.
\end{thm}

In view of Theorem \ref{limit} and since $\kappa\geq 0$, and clearly satisfies $\kappa=0$ only if  ${\cal{H}}_{\bb{\vartheta}}$ is true, it is reasonable to reject the null hypothesis for large values of the test statistic $\widehat T^{(\Psi)}_n $, as intuitively stated in the previous section.

\subsection{Weak limit of the test statistic} \label{part.two}

In the next theorem we study the asymptotic distribution of the test statistic $\widehat T^{(\Psi)}_{n}$ under the null hypothesis ${\cal{H}}_{\bb{\vartheta}}$. In what follows, $ \overset{\mathbb{D}} {\to}$ denotes convergence in distribution and $\circ$ denotes the Hadamard product.


\begin{thm} \label{asympt.null.distrib}
Let $\bb{X}_1, \ldots, \bb{X}_n$ be i.i.d. copies of the random variable $\bb{X} \in \mathbb{R}^p$. Then under the conditions \eqref{equiv}--\eqref{bahadur2}, 
and provided that $\int \|\bb{t}\|^4w(\bb{t}){\rm{d}}\bb{t}<\infty$, we have under the null hypothesis  ${\cal{H}}_{\bb{\vartheta}}$ figuring in \eqref{null2} that,
$$ \widehat T^{(\Psi)}_{n} \ \overset{\mathbb{D}} {\to} \ \|Z_1+Z_2\|_w^2 -2,$$ where $Z_1$ and $Z_2$ are two independent centered Gaussian random elements of $L^2_w$ having covariance kernels
\begin{eqnarray*}
K_1(\bb{t},\bb{s}) & = & \mathbb{E}\Big[W_1(\bb{X}_0, \bb{V}, \Psi_0;\bb{t})W_1(\bb{X}_0, \bb{V}, \Psi_0; \bb{s})\Big],\\
K_2(\bb{t},\bb{s}) & = & \mathbb{E}\Big[W_2(\bb{X}_0, \Psi_0; \bb{t}) W_2(\bb{X}_0, \Psi_0; \bb{s})\Big],
\end{eqnarray*}
respectively, where
\begin{eqnarray*}
W_1(\bb{X},\bb{V}, \Psi_0; \bb{t}) & = & -2\Psi'_0(\|\bb{t}\|^2) \bb{t}^\top {\bb{L}}_0(\bb{X})\circ \bb{V}^{-1/2} \bb{\Lambda}  \bb{t} -\Psi_0(\|\bb{t}\|^2) \bb{t}^\top {\bb{\ell}}_0(\bb{X})+W_2(\bb{X}, \Psi_0; \bb{t}),\\
W_2(\bb{X}, \Psi_0; \bb{t}) & = & \cos(\bb{t}^\top \bb{X})-\Psi_0(\|\bb{t}\|^2)+\sin(\bb{t}^\top \bb{X}),
\end{eqnarray*}
and the matrix $\bb{\Lambda}$ is defined in the statement of Lemma \ref{raiz.de.S} in the Appendix.
\end{thm}

As an immediate consequence of Theorem \ref{limit} and Theorem \ref{asympt.null.distrib}, it follows that the test which rejects the null hypothesis ${\cal{H}}_{\bb{\vartheta}}$  for large values of $\widehat T^{(\Psi)}_{n}$ is consistent against each fixed alternative distribution.

Clearly the covariance kernel $K_2$  does not depend on the scatter matrix $\bb{V}$. The next results shows that, under some mild assumptions,  $K_1$ also does not involve  $\bb{V}$.
Recall that $\bb{L}_0(\cdot)=(\bb{L}_0(\cdot)_{rs})_{1\leq r,s\leq p}$ stands for $\bb{L}_{\bb{\vartheta}}(\cdot)$ when  $\bb{\vartheta}=\bb{\vartheta}_0$.

\begin{prop} \label{la.cova}
Suppose that the assumptions in Theorem \ref{asympt.null.distrib} hold and that for some scalar functions $\alpha(\cdot)$ and $\beta(\cdot)$ such that $\alpha(\bb{t})=\alpha(\|\bb{t}\|)$ and $\beta(\bb{t})=\beta(\|\bb{t}\|)$ we can write \begin{equation} \label{condition}
\mathbb{E}\Big[\cos(\bb{t}^\top X_0)\bb{L}_0(X_0)\Big]=\alpha(\bb{t}){\rm I}_p+\beta(\bb{t})\bb{t}\bb{t}^\top.
\end{equation}
Then, the covariance kernels $K_1$ and $K_2$ may be written as
\begin{eqnarray*}
K_1(\bb{t},\bb{s}) & = & \Psi_0(\|\bb{t}\|^2)\Psi_0(\|\bb{s}\|^2)\bb{t}^\top \mathbb{V}\{\bb{\ell}_0(\bb{X}_0)\}\bb{s}\\
         & &    -\Psi_0(\|\bb{t}\|^2)\bb{t}^\top\mathbb{E}\{\bb{\ell}_0(\bb{X}_0)
         \sin(\bb{s}^\top \bb{X}_0) \} -\Psi_0(\|\bb{s}\|^2)\bb{s}^\top\mathbb{E}\{\bb{\ell}_0(\bb{X}_0)\sin(\bb{t}^\top \bb{X}_0)\}\\
        & &     -\Psi_0'(\|\bb{t}\|^2)\left\{\alpha(\bb{s})\|\bb{t}\|^2+\beta(\bb{s})(\bb{t}^\top \bb{s})^2 \right\}
                -\Psi_0'(\|\bb{t}\|^2)\left\{\alpha(\bb{t})\|\bb{s}\|^2+\beta(\bb{t}) (\bb{t}^\top \bb{s})^2 \right\}\\
         & &    +\Psi_0'(\|\bb{t}\|^2)\Psi_0'(\|\bb{s}\|^2)\left\{ 2\sigma_1 (\bb{t}^\top \bb{s})^2+\sigma_2\| \bb{t}\|^2\|\bb{s}\|^2\right\}+K_2(\bb{t},\bb{s}),\\
K_2(\bb{t},\bb{s}) & = & \Psi_0(\|\bb{t}-\bb{s}\|^2)-\Psi_0(\|\bb{t}\|^2)\Psi_0(\|\bb{s}\|^2),
\end{eqnarray*}
where $\sigma_1=\mathbb{E}\left\{ \bb{L}_0(\bb{X}_0)_{12}^2 \right\}$ and $\sigma_2=\mathbb{E}\left\{ \bb{L}_0(\bb{X}_0)_{11}\bb{L}_0(\bb{X}_0)_{22} \right\}$.
\end{prop}
Proposition \ref{la.cova} implies that the limit null distribution of the test statistic $\widehat T^{(\Psi)}_{n}$ depends on $\Psi_{0}(\cdot)$ and the type of estimator used in estimating $\bb{\vartheta}$, but does not involve neither the location $\bb{\delta}$ nor the scatter matrix $\bb{V}$. Thus the test based on  $\widehat T^{(\Psi)}_{n}$ is asymptotically invariant--in--law.

We note that \eqref{condition} is fulfilled when $\bb{V}$ is estimated with the ME. To see this recall that the ME of $\bb{V}$ is given by  $\widehat {\bb{V}}_n=\{-2\Psi_0'(0)\}^{-1}\bb{S}_n$. Thereby and ignoring the factor $\{-2\Psi_0'(0)\}^{-1}$, we have that $\bb{L}_0(\bb{X})=\bb{X}\bb{X}^\top-{\rm I}_p$, and thus the specification of the functions figuring in   \eqref{condition} under ME is $\alpha(\bb{t})=-\left\{2\Psi_0'(\|\bb{t}\|^2)+\Psi_0(\|\bb{t}\|^2)\right\}$ and $\beta(\bb{t})=-4\Psi_0''(\|\bb{t}\|^2)$.
 It can be  checked that \eqref{condition} is  also fulfilled when $\bb{V}$ is estimated by the MLE (see Chapter 13 of \citealp{Bilodeau1999}).

\begin{rem}
Despite the fact that the test statistic $\widehat T^{(\Psi)}_{n}$ is asymptotically invariant--in--law, $\widehat T^{(\Psi)}_{n}$ is not affine invariant. To see this, write $\widehat T^{(\Psi)}_{n}({\cal{X}}_n;{\cal{X}}_{0,n})$ for the test statistic based on ${\cal{X}}_{n}=(\bb{X}_{1}, \ldots, \bb{X}_{n})$ and ${\cal{X}}_{0,n}=(\bb{X}_{0,1}, \ldots, \bb{X}_{0,n})$, and let $\bb{A}$ be a full--rank matrix of dimension $(p\times p)$. 
Then, it can be  easily checked that
\begin{equation} \label{ne}
\widehat T^{(\Psi)}_{n}(\bb{A}{\cal{X}}_n; {\cal{X}}_{0,n})= \widehat T^{(\Psi)}_{n}({\cal{X}}_n; \bb{U}{\cal {X}}_{0,n}),
\end{equation}
where $\bb{U} \in \mathcal{O}_p$ is an orthogonal matrix defined by $\bb{U}=\bb{A}^\top \widehat{\bb{V}}_n^{1/2}(\bb{A} \widehat{\bb{V}}_n \bb{A}^\top)^{-1/2}$. This shows that, in general, $\widehat T^{(\Psi)}_{n}(\bb{A}{\cal{X}}_n; {\cal{X}}_{0,n}) \neq \widehat T^{(\Psi)}_{n}({\cal{X}}_n, {\cal{X}}_{0,n})$. Specifically, equation \eqref{ne}  shows that the transformation $\bb{X}_j \mapsto \bb{A}\bb{X}_j, \ j=1,\ldots, n$, entails a rotation of the sample from $X_0$, to a sample from $OX_0$, where the rotation depends on the data
${\cal{X}}_n$ as well as on the matrix $A$.  Since however ${\cal{X}}_{0,n}$ is a random sample from a spherical distribution and since $\bb{U}$ is an orthogonal matrix it follows that $\bb{U}{\cal{X}}_{0,n}$ is also a sample from the same distribution.
\end{rem}

\subsection{Asymptotic power of the test} \label{part.three}

This subsection is devoted to derive the asymptotic distribution  of the test statistic $\widehat T^{(\Psi)}_{n}$ under fixed alternatives.
To this end we refer to the quantity $\kappa$ figuring in Theorem \ref{limit} and analogously to \eqref{T2} and \eqref{Gn}, and write  $\kappa=\|G\|_w^2>0$, where
\begin{equation*}
G(\bb{t})={\rm Re}(\varphi_{Y}(\bb{t})) + {\rm Im}(\varphi_{Y}(\bb{t})) - {\rm Re}(\varphi_{0}(\bb{t})) - {\rm Im}(\varphi_{0}(\bb{t})),
\end{equation*}
with $\bb{Y}=\bb{V}^{-1/2}\bb{X}$.

\begin{thm} \label{power}
Let $\bb{X}_1, \ldots, \bb{X}_n$ be i.i.d. copies of the random variable $\bb{X} \in \mathbb{R}^p$, such that $\mathbb E (\|\bb{X}\|^2)<\infty$ and $\kappa>0$, write $ \bb{V}={\rm{diag}}(\lambda_1, \ldots, \lambda_p)$, and assume that $\displaystyle \min_{1\leq j\leq p} \lambda_j>0$. Then under conditions \eqref{equiv}--\eqref{bahadur2},
and provided that $\int \|\bb{t}\|^4w(\bb{t})d\bb{t}<\infty$, we have
\[
\sqrt{n}\left(\frac{\widehat T^{(\Psi)}_{n}}{n}-\kappa\right) \ \overset{\mathbb{D}}{\to} \ {\cal{N}}(0,\sigma_1^2+\sigma^2_2),
\]
where
\[
\sigma_1^2=\int \int K^{({\rm{a}})}_1(\bb{t},\bb{s}) G(\bb{t})G(\bb{s})w(\bb{t})w(\bb{s})d\bb{t}d\bb{s},
\]
\[
\sigma_2^2=\int \int K_2(\bb{t},\bb{s})G(\bb{t})G(\bb{s})w(\bb{t})w(\bb{s})d\bb{t}d\bb{s},
\]
\[
K^{({\rm{a}})}_1(\bb{t},\bb{s}) =\mathbb{E}\left [ W^{({\rm{a}})}_1(\bb{Y}; \bb{t}) W^{({\rm{a}})}_1(\bb{Y}; \bb{s}) \right],
\]
\[W^{({\rm{a}})}_1(\bb{Y}; \bb{t})  =  C_{\bb{Y}}(\bb{t})+\cos(\bb{t}^\top \bb{Y})-{\rm Re}( \varphi_{\bb{Y}}(\bb{t}))
 +S_{\bb{Y}}(\bb{t})+\sin(\bb{t}^\top \bb{Y})-{\rm Im}( \varphi_Y(\bb{t})),
 \]
 $C_{\bb{Y}}(\cdot)$ and $S_{\bb{Y}}(\cdot)$ are as defined in Lemma \ref{lemma3} in the Appendix and $\bb{Y}=\bb{V}^{-1/2}\bb{X}$.
\end{thm}

Theorem \ref{power} allows to approximate the power of the test based on   $\widehat T^{(\Psi)}_{n}$. To see this, let
$c_{n,\alpha}$ denote the upper $\alpha$-percentile of the null distribution of $\widehat T^{(\Psi)}_{n}$.  Then from Theorem \ref{asympt.null.distrib} and Theorem \ref{power} we have,
\begin{eqnarray}
\mathbb P(\widehat{T}^{(\Psi)}_{n} > c_{n,\alpha}) & = & \mathbb P\left\{ \sqrt{n}\left(\frac{\widehat T^{(\Psi)}_{n}}{n}-\kappa \right) > \sqrt{n}\left(\frac{c_{n,\alpha}}{n}-\kappa\right) \right\} \nonumber \\
 & \approx  & \Phi \left(\frac{\sqrt{n}\kappa}{\sqrt{\sigma^2_1+\sigma_2^2}}\right) \label{pw1}
\end{eqnarray}
where $\Phi(\cdot)$ is the distribution function of the standard normal law.

\begin{rem} \label{zero} The most popular CF--based statistic for goodness--of--fit is the BHEP test for testing the composite null hypothesis of multivariate normality; see \cite{Epps1983} and \cite{Henze1997}. This test, analogously to \eqref{T2}, may be written as
\begin{equation} \nonumber
\widehat T^{(G)}_{n}=n\| \varphi_n-\varphi_0\|_w^2,
\end{equation}
with $\varphi_0(\bb{t})=e^{-\|{\bb{t}}\|^2/2}$ and $w(\cdot)$ being the density of the $p$--variate normal law ${\cal{N}}(\bb{0},(1/\beta^2){\rm{I}}_p), \ \beta>0$. In the BHEP test,  the empirical CF $\varphi_n(\cdot)$ is computed as in \eqref{ecf} using the standardized observations in \eqref{stand} with  $\widehat{\bb{\delta}}_n=\overline{\bb{X}}_n$ and $\widehat{\bb{V}}_n=\bb{S}_n$. Under the null hypothesis, $\widehat T^{(G)}_{n}$ converges in law to $\|Z_1\|_w^2$, where $Z_1$ is as in Theorem  \ref{asympt.null.distrib}. The expression of $K_1(\bb{t},\bb{s})$   in Proposition \ref{la.cova} coincides with that given in \cite{Henze1997}.
The behaviour of this statistic under alternatives has been studied by \cite{Baringhaus2017}. It specifically follows, in analogy to \eqref{pw1},  that if
$c^{(G)}_{n,\alpha}$ denotes the upper $\alpha$--percentile of the null distribution of $\widehat T^{(G)}_{n}$, then
\begin{equation} \label{pw2}
\mathbb P(\widehat T^{(G)}_{n}> c^{(G)}_{n,\alpha})\approx \Phi \left(\frac{\sqrt{n}\kappa}{\sigma_1}\right).
\end{equation}
From \eqref{pw1} and \eqref{pw2} it follows that the corresponding tests, are both consistent against alternatives with $\kappa>0$. However for moderate sample sizes  the BHEP test is expected to be more powerful than the test based on $\widehat{T}^{(\Psi)}_{n}$, since the latter also involves the extra factor $\sigma_2$ corresponding to the randomness introduced by ${\cal{X}}_{0,n}$, although the difference in their powers should decrease as the sample size $n$ increases. This fact is corroborated by the simulation results of Section \ref{sec7}. The drawback with the BHEP test is that the formulation of this test, although in principle it does carry  over to alternative families of distributions (besides the Gaussian distributions),  it does not produce a test statistic that can be straightforwardly  computed as that corresponding to the new test figuring in \eqref{ts1}.
\end{rem}


\section{Computational procedure for test implementation}\label{sec6}

This section addresses two issues related to the practical application of the proposed test of  ${\cal{H}}_{\bb{\vartheta}}$ in \eqref{null2}, namely, the randomness of the test due to $\bb{X}_{0,1}, \ldots, \bb{X}_{0,n}$ and the calculation of the critical point.

\subsection{Dealing with randomness}
By construction of the test statistic  $\widehat{T}^{(\Psi)}_{n}$ in \eqref{ts1},
for fixed data ${\cal{X}}_n=(\bb{X}_1, \ldots, \bb{X}_n)$, the decision of rejecting or not rejecting the null hypothesis ${\cal{H}}_{\bb{\vartheta}}$ depends on the sample ${\cal{X}}_{0,n}=(\bb{X}_{0,1},\ldots,\bb{X}_{0,n})$ generated from $\bb{X}_0$. This drawback is encountered by other procedures such as  the random projection procedure; see, for example, \cite{Cuesta2006}. The following suggestion for resolving this randomness, inherent in the proposed test statistic,
borrows from analogous resolutions in the case of random projection methods. Specifically, we generate $m$ samples from $\bb{X}_0$, say ${\cal{X}}_{0,n}^{r}$,  $r=1,\ldots,m$, and compute corresponding test statistics
$\widehat T^{(\Psi),r}_{n} =\widehat T^{(\Psi)}_{n} ({\cal{X}}_n;{\cal{X}}_{0,n}^{r})$, $r=1,\ldots,m$.  Then we suggest to take as ``final" test statistic  the maximum or the sample mean of the test statistics  $\widehat T^{(\Psi)}_{n} ({\cal{X}}_n;{\cal{X}}_{0,n}^{r})$, $r=1,\ldots,m$, so computed. The choice of $m$ will be numerically studied in Section \ref{sec7} by means of a battery of simulation experiments.

\subsection{Approximating  critical points}

As mentioned in Section \ref{sec5}, when the unknown parameter $\bb{\vartheta}$ is estimated by ME or MLE, the asymptotic null distribution of the proposed test statistic $\widehat T^{(\Psi)}_{n}$ in \eqref{ts1} does not depend on $\bb{\vartheta}$. As a consequence, the 
null distribution can be consistently estimated acting as if $\bb{\vartheta}=\bb{\vartheta}_0$, where $\bb{\vartheta}_0=(\bb{0},\rm{I}_p)$.
Specifically, 
the test based on the sample mean (and analogously for the test based on the maximum) is carried out as follows:
\begin{enumerate}[Step 1.]
  \item (Original statistic) Generate $m$ samples ${\cal{X}}_{0,n}^r, \ r=1,\ldots,m$, under the null hypothesis ${\cal{H}}_{\bb{\vartheta}_0}$,  and for each sample compute the corresponding value $\widehat T^{(\Psi),r}_{n} = \widehat T^{(\Psi)}_{n} ({\cal{X}}_n; {\cal{X}}_{0,n}^r)$ according to \eqref{ts1}, and then set as test statistic the quantity
      \begin{equation} \label{TnMEAN}
        \overline {\widehat T}_{n,m}^{(\Psi)} = \frac{1}{m} \sum_{r=1}^m \widehat T^{(\Psi),r}_{n}.
      \end{equation}
  \item (Re--sample statistic) Generate $m+1$ re--samples ${\cal{X}}_{0,1}^{r*},  \ r=1,\ldots,m+1$, under the null hypothesis ${\cal{H}}_{\bb{\vartheta}_0}$, and compute
      $\overline {\widehat T}_{n,m}^{(\Psi),*} = m^{-1} \sum_{r=1}^m \widehat T^{(\Psi)}_{n} ({\cal{X}}_{0,n}^{m+1*}; {\cal{X}}_{0,n}^{r*})$.
  \item (Critical point) Repeat Step 2 a number of times, say $M$, and thereby compute the $(1-\alpha)\%$ quantile $c_{n,\alpha}$ of the empirical distribution of  the set $\Big\{\overline {\widehat T}_{n,m,j}^{(\Psi),*}\Big\}_{j=1}^M$.
  \item (Decision) Reject the null hypothesis ${\cal{H}}_{\bb{\vartheta}}$ figuring in \eqref{null2} if $ \overline {\widehat T}_{n,m}^{(\Psi)} > c_{n,\alpha}$, where $\overline {\widehat T}_{n,m}^{(\Psi)}$ is the original test statistic computed in Step 1.
  \end{enumerate}

\begin{rem} Since, roughly speaking, the proposed approach
consists in transforming the goodness--of--fit problem into a two-sample problem, other approaches commonly used to approximate the null distribution of statistics for the latter problem could be used. Specifically, nonparametric bootstrap, permutation and
weighted bootstrap approximations are studied in \cite{Meintanis2005}, \cite{Henze2005}, \cite{Alba2008}, \cite{Alba2017}, \cite{JG2017} and \cite{Chen2019}  in order to approximate the null distribution of  statistics, which are similar to the one proposed in this paper, for the  two-sample problem. Although those procedures, could potentially also be used  to estimate $c_{n,\alpha}$, we feel that the above proposed approximation, which is a version of the parametric bootstrap, is more natural to apply in the current setting of goodness--of--fit testing for parametric distributions. 
\end{rem}

\section{Numerical studies} \label{sec7}
In Subsection \ref{MC} we include the results of a fairly extensive simulation study of the behavior of the new test as compared to alternative procedures in a number of sampling situations. A real--data application illustrates the use of the new test with actual data in Subsection \ref{realdata}.
The \textsf{R} code for all the numerical studies is provided as Electric Supplementary Material.

\subsection{Simulations} \label{MC}
In this subsection, a series of simulation experiments is carried out in order to assess the finite--sample performance of the proposed test statistic $\widehat T^{(\Psi)}_{n}$ in \eqref{ts1}.
Specifically, goodness--of--fit tests for four well-known elliptical distributions, the multivariate normal distribution, the multivariate Laplace distribution, the multivariate Student--t distribution, and the Kotz--type distribution, are studied.

In the following, we employ the multivariate standard normal density as the weight function, which means that the kernel $\Psi(\xi)=e^{-\xi/2}$ (a slight variation of the spherical stable kernel $e^{-\xi^{b/2}}$ for $b=2$) is applied on $\widehat T^{(\Psi)}_{n}$ in \eqref{ts1}, and denote by $\overline {\widehat T}_{n,m}^{({\cal{N}})}$ the test corresponding to \eqref{TnMEAN} and by $\widetilde {\widehat  T}_{n,m}^{({\cal{N}})}$, the analogous test where in \eqref{TnMEAN} instead of the mean we employ the maximum of $\widehat T^{(\Psi),r}_{n}$, $r=1,\ldots,m$.
Furthermore, all tests are performed at the $5\%$ significance level using $1000$ trails with $M=1000$ in Step 3 above.
The dimensions of ${\bb X}$ considered are $p \in \{2,3,5\}$, and the sample sizes are $n \in \{20,50,100\}$.

For comparison purposes, we consider the tests in \cite{Ducharme2020} and \cite{Hallin2021} as benchmarks, both of which are applicable for elliptical distributions.
\cite{Hallin2021}'s test is based on the Wasserstein distance between a discrete empirical distribution and a continuous distribution specified by the null hypothesis. Sophisticated numerical techniques are involved in computing this test statistic. On the other hand, \cite{Ducharme2020}'s smooth test is relatively easier to apply, but the global consistency of this test can only be obtained when the value of the hyperparameter increases.
In the following, we represent the test statistic in \cite{Ducharme2020} with $DM$ and choose the hyperparameter $K=7$ as suggested in the \textsf{R} package \texttt{ECGofTestDx} (\citealp{ECGofTestDx}), which is available on the CRAN and for the implementation of $DM$.
The test statistics based on the $1$- and $2$-Wasserstein distance in \cite{Hallin2021} are denoted by $HMS_1$ and $HMS_2$, respectively.
For the implementation of $HMS_1$ and $HMS_2$ we refer to \texttt{https://github.com/gmordant/WassersteinGoF}. Moreover and since in the case of testing for normality the new test statistic is essentially equivalent to the BHEP test with $\beta=1$ ($BHEP_1$), we compare the performance of the proposed tests $\overline {\widehat T}_{n,m}^{({\cal{N}})}$ and $\widetilde {\widehat T}_{n,m}^{({\cal{N}})}$ with that of $BHEP_1$ in order to evaluate the effect of the artificial data. In the case of testing for the multivariate Laplace distribution we  compare our test with the test of \cite{Fragiadakis2011} which is specifically tailored for goodness--of--fit test for this distribution and is also based on CFs.
For the numerical implementations of these alternative methods, critical points of the $DM$ test can be computed via an asymptotic chi-square approximation as shown in the \textsf{R} package \texttt{ECGofTestDx}.
For other tests, critical points are approximated via a Monte Carlo algorithm.
Specifically, we draw a large number of independent random samples, say $B$, according to the null hypothesis, calculate the test statistic for each such sample, and calculate the empirical quantile of the $B$ simulated test statistics as the critical point.
In contrast to Section 6.2, this algorithm does not require the $m$ repeat resamples needed by the proposed tests due to the inherent randomness.
The number of resamples was set to $B=1000$, analogously to the resample size $M=1000$ employed for the new test.

\textbf{Choice of the Monte Carlo sample size:} First, we investigate numerically the impact of the Monte Carlo size $m$ in Step 1,  on both the test statistic $\overline {\widehat T}_{n,m}^{(\Psi)}$  in \eqref{TnMEAN}, as well as on the test based on $\widetilde {\widehat T}_{n,m}^{(\Psi)}$ (always with $\Psi(\xi)=e^{-\xi/2}$).
With this aim, we calculate the percentage of rejection of the null hypothesis among the $1000$ replicates with $m \in \{1,5,10,15,20,25,30\}$.
We consider the goodness--of--fit test for multivariate normality as an example and take four cases of the distributions considered by \cite{Ebner2020}. Specifically, these distributions are $\mathcal{N}_p(\bb{e}_p,\bb{ \Sigma}_{0.5})$, $\mathcal{NM}_p(3)$, $\mathcal{U}^p(0,1)$, and $\mathcal{MAR}_p(Exp)$, Case 1--4 respectively, whereby $\mathcal{N}_p(\bb{\mu},\bb{\Sigma})$ stands for a $p$-dimensional Gaussian distribution with mean vector $\bb{\mu}$ and covariance matrix $\bb{\Sigma}$, $\bb{ e}_p=(1,2,\ldots,p)^\top$, and $\bb{\Sigma}_{0.5}$ represents a positive definite $(p\times p)$ matrix with diagonal elements equal to 1 and off-diagonal elements equal to 0.5. In this connection,  Case 1 is considered in order to show that the new tests are asymptotically invariant--in--law, i.e. that indeed their Type--I error is equal to the nominal level of $5\%$. As for alternatives, in Case 2, $\mathcal{NM}_p(\theta)$ denotes a balanced mixture of the standard Gaussian distribution and $\mathcal{N}_p(\theta {\bb{1}}_p,{\rm{I}}_p)$,  where ${\bb{1}}_p=(1,1,\ldots,1)^\top$, in Case 3, $\mathcal{U}^p(0,1)$ denotes a $p$-dimensional random vector whose coordinates are i.i.d. copies of univariate uniform distribution in the interval $(0,1)$, and finally $\mathcal{MAR}_p(Exp)$ denotes a $p$-dimensional standard Gaussian distribution whose $p$th component is independently replaced by an observation from a unit exponential distribution (Case 4). As for the unknown parameters figuring in ${\cal{H}}_{\bb{\vartheta}}$ when testing for normality, we estimate the mean vector and covariance matrix by the sample mean $\overline {\bb{X}}_n$ and sample covariance matrix $\bb{S}_n$, respectively.

\begin{figure}[!htb]
 \centering
 \includegraphics [scale=0.5]{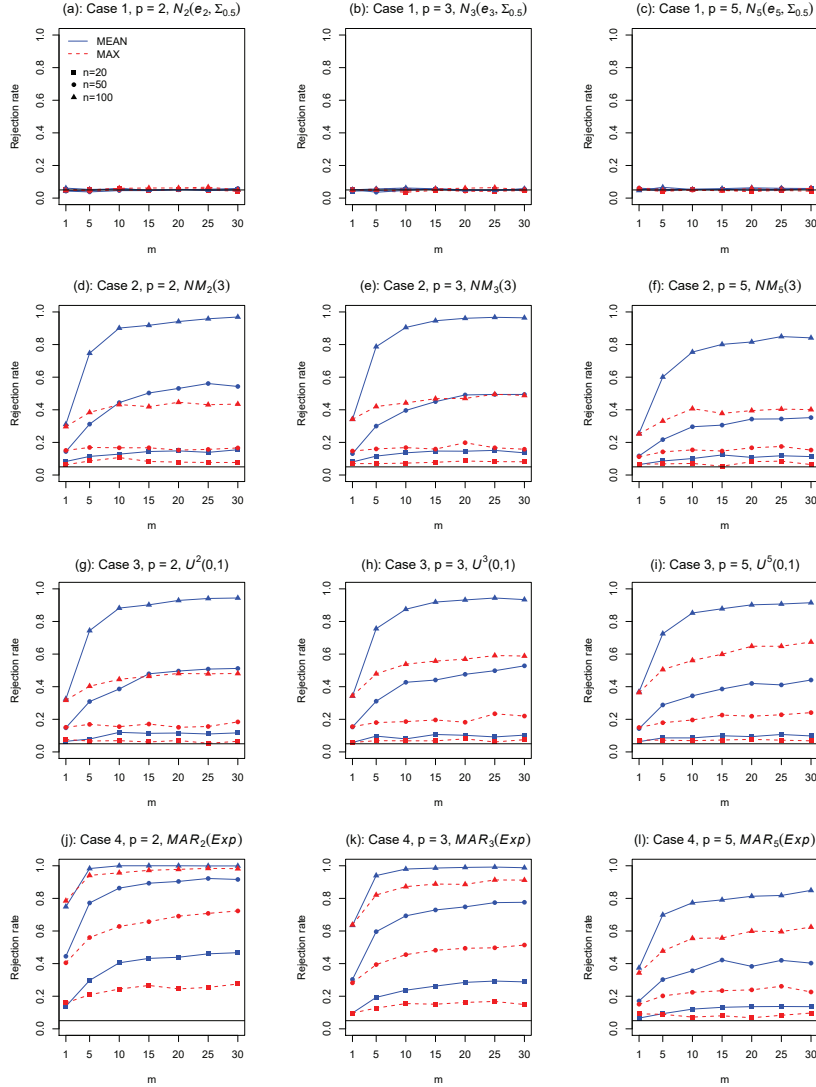}\par
 \vspace{-0.5cm}
 \caption{Empirical rejection rates of $\overline {\widehat T}_{n,m}^{({\cal{N}})}$ (MEAN) and $\widetilde {\widehat T}_{n,m}^{({\cal{N}})}$ (MAX) against different values of $m$ for Cases 1-4. The horizontal line corresponds to the $5\%$ significance level. }  \label{fig-m}
\end{figure}

Figure \ref{fig-m} displays the empirical rejection rates corresponding to $\overline {\widehat T}_{n,m}^{({\cal{N}})}$ and $\widetilde {\widehat T}_{n,m}^{({\cal{N}})}$ against different values of $m$ at the $5\%$ significance level for each case of distribution.
It can be seen that the Type--I error rates (Case 1) can be controlled well for all the choices of $m$.
As for the empirical power (Cases 2--4), generally, it increases at first as $m$ increases, while as $m$ continues to grow, the empirical power tends to be stable.
Thus in the following we set $m=10$, in order to strike a balance between higher empirical power and lower computation time.

\noindent
\textbf{Example 1: Tests for the multivariate Normal distribution}

We generate observations from $\mathcal{MT}_p(\bb{0}, {\rm{I}}_p, \nu)$ with $\nu \in \{1,3,5,10,15,25,\inf\}$. Here, $\mathcal{MT}_p(\bb{\mu}, \bb{\Sigma}, \nu)$ denotes the multivariate Student--t distribution (Kotz and Nadarajah, 2004) with location parameter $\bb{\mu}$, scale matrix $\bb{\Sigma}$ and degrees of freedom $\nu$, so that $\mathcal{MT}_p(\bb{0}, {\rm{I}}_p, \inf)$ coincides with the normal distribution.

\begin{figure}[!htb]
 \centering
 \includegraphics [scale=0.54]{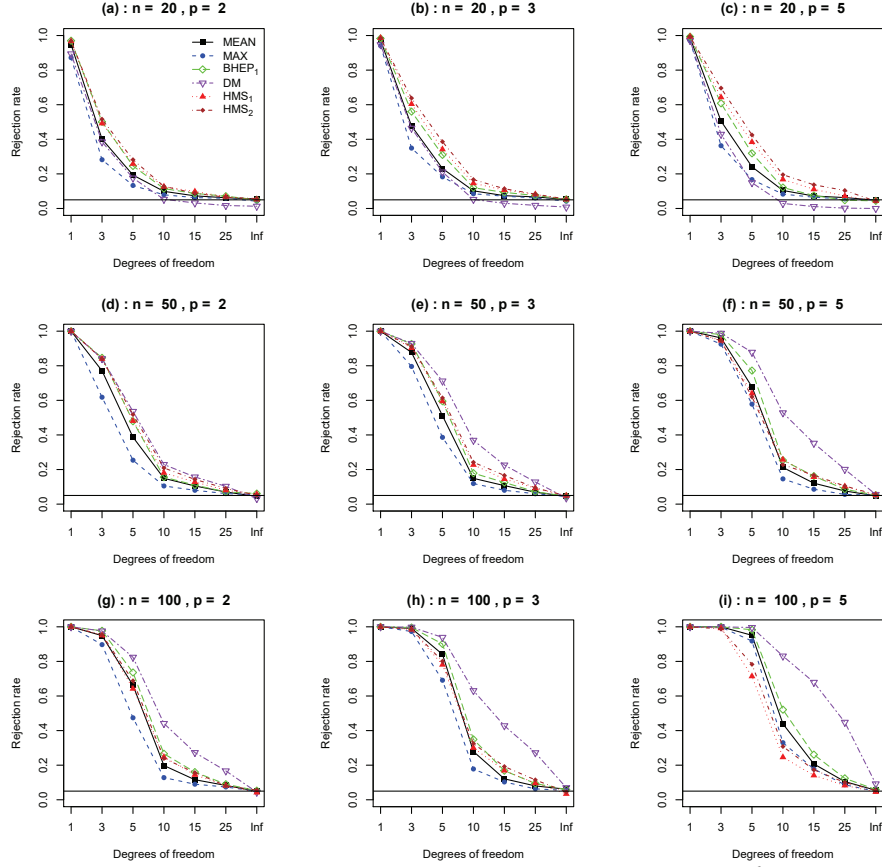}\par
 \vspace{-0.5cm}
 \caption{Simulation results for Example 1. Empirical rejection rates of $\overline {\widehat T}_{n,10}^{({\cal{N}})}$ (MEAN), $\widetilde {\widehat T}_{n,10}^{({\cal{N}})}$ (MAX), $BHEP_1$, $DM$, $HMS_1$, and $HMS_2$ against the degrees of freedom $\nu$. The horizontal line corresponds to the $5\%$ significance level.}  \label{fig-mnt}
\end{figure}

Now we calculate the empirical sizes and powers of $\overline {\widehat T}_{n,10}^{({\cal{N}})}$, $\widetilde {\widehat T}_{n,10}^{({\cal{N}})}$, $BHEP_1$, $DM$, $HMS_1$, and $HMS_2$, and plot the rejection frequencies against the degrees of freedom $\nu$ in Figure \ref{fig-mnt}.
We can observe that for the null hypothesis ($\nu=\inf$), the Type--I error rates can be controlled well for all scenarios except for the $DM$ test. It can also be observed that with sample size $n$ increasing and/or the degrees of freedom $\nu$ decreasing, the empirical powers of all the tests increase, which verifies the consistency of these methods. In general, the performance of $\overline {\widehat T}_{n,10}^{({\cal{N}})}$, $BHEP_1$, $HMS_1$, and $HMS_2$ are comparable in this example, with the test based on $\widetilde {\widehat T}_{n,10}^{({\cal{N}})}$ being less competitive. As for the test $DM$, though it is the most powerful when $n=50$ and $n=100$, its empirical rejection rates are the lowest for sample size $n=20$. Furthermore, the empirical powers of the proposed test $\overline {\widehat T}_{n,10}^{({\cal{N}})}$ is slightly lower than that of  $BHEP_1$ for all scenarios. Hence, it appears that the tests with artificial data are less powerful than tests where expectations can be explicitly calculated rather than simulated. This finding corroborates the theoretical results in Section \ref{sec5} which imply that the randomness due to the artificial data increases the variance of the Gaussian process involved in the limit distribution.

\noindent
\textbf{Example 2: Tests for the  multivariate Laplace distribution}

In this example, we apply the proposed tests to testing goodness--of--fit  for the multivariate Laplace distribution $\mathcal{ML}_p(\bb{\delta},\bb{V})$ (Fang et al. 1990), where $\bb{\delta}$ and $\bb{V}$ denote the mean vector and covariance matrix, respectively.
Then we estimate the parameters $\bb{\delta}$ and $\bb{V}$ by the MEs ${\widehat {\bb{\delta}}}_n = \overline {\bb{X}}_n$ and $\widehat {\bb{V}}_n=\bb{S}_n$, respectively.
The  multivariate Laplace distribution is an attractive alternative to the multivariate Gaussian distribution due to its heavier tails, and can be generated with the function \emph{rmvl} from the \textsf{R} package \texttt{LaplacesDemon} (\citealp{LaplacesDemon}).
For the Gaussian weight function $e^{-a\|\cdot\|^2}$, $a>0$, $FM_a$, stands for the test of \cite{Fragiadakis2011} which may be rendered in a simple closed formula. In this example $FM_1$, $HMS_1$, and $HMS_2$ are employed as competitor tests.

Observations are generated from a mixture of a standard Laplace with a standard Gaussian distribution $(1-\theta)\ \mathcal{ML}_p(\bb{0},{\rm{I}}_p) + \theta\ \mathcal{N}_p(\bb{0},{\rm{I}}_p)$, where $\theta$ denotes the mixture parameter. We choose $\theta \in \{0,0.2,0.4,0.6,0.8,1\}$, with $\theta=0$ corresponding to the null hypothesis, while $\theta>0$ corresponds to alternatives.

\begin{figure}[!htb]
 \centering
 \includegraphics [scale=0.54]{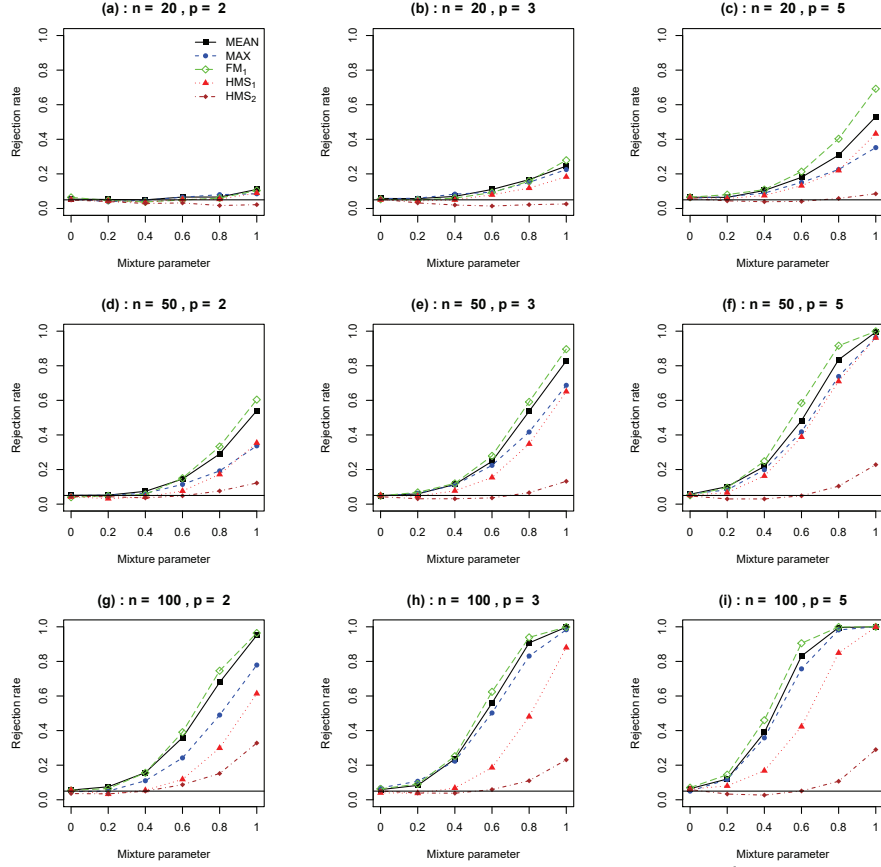}\par
 \vspace{-0.5cm}
 \caption{Simulation results for Example 2. Empirical rejection rates of $\overline {\widehat T}_{n,10}^{({\cal{N}})}$ (MEAN), $\widetilde {\widehat T}_{n,10}^{({\cal{N}})}$ (MAX), $FM_1$, $HMS_1$, and $HMS_2$ against the mixture parameter $\theta$. The horizontal line corresponds to the $5\%$ significance level.}  \label{fig-mlt}
\end{figure}

The rejection rates for $\overline {\widehat T}_{n,10}^{({\cal{N}})}$, $\widetilde {\widehat T}_{n,10}^{({\cal{N}})}$, $FM_1$, $HMS_1$, and $HMS_2$, are plotted against the mixture parameter $\theta$ in Figure \ref{fig-mlt}.
It can be observed that for the null hypothesis ($\theta=0$), the Type--I error rates can be controlled well for all scenarios.
For the alternatives ($\theta>0$), we can observe that the empirical powers of all tests except $HMS_2$ increase with increasing sample size $n$ and/or mixture parameter $\theta$, which again is in line with the consistency results.
However, the empirical powers of $HMS_2$ are lower than the nominal level for smaller sample sizes ($n=20, 50$) or mixture parameters ($\theta\leq 0.4$), which means that this test fails to detect the alternatives.
Figure \ref{fig-mlt} also indicates that $FM_1$ and the proposed tests generally outperform $HMS_1$ and $HMS_2$. Now $FM_1$ performs slightly better than the proposed tests  $\overline {\widehat T}_{n,10}^{({\cal{N}})}$ and  $\widetilde {\widehat T}_{n,10}^{({\cal{N}})}$ in most cases.
Recall however that the test based on $FM_a$ is tailored for the multivariate Laplace distribution, and can be explicitly calculated. In other words we encounter here the same kind of behaviour as in the case of testing for the normal distribution and the $BHEP$ test.

\noindent
\textbf{Example 3: Tests for the multivariate Student--t distribution}

In this example, we apply the proposed tests to testing goodness--of--fit  for the
multivariate Student--t distribution $\mathcal{MT}_p(\bb{\delta}, \bb{V}, \nu)$ with known degrees of freedom $\nu$.
In this connection recall that if $\bb{X}\sim \mathcal{MT}_p(\bb{\delta}, \bb{V}, \nu)$, then $\mathbb{E}(\bb{X})=\bb{\delta}$ for $\nu >1$ and $\mathbb{V}(\bb{X})=(\nu/(\nu-2))\bb{V}$ for $\nu>2$. Thus we focus on the case where $\nu>2$ and estimate the parameters $\bb{\delta}$ and $\bb{V}$ by the MEs ${\widehat {\bb{\delta}}}_n = \overline {\bb{X}}_n$ and $\widehat {\bb{V}}_n=((\nu-2)/\nu) \bb{S}_n$, respectively.

Specifically, and similar to Section 5.2.2 in \cite{Hallin2021}, we apply the test for the multivariate Student--t distribution $\mathcal{MT}_p(\bb{0}, {\rm{I}}_p, \nu)$ with $\nu=12$,  against multivariate skew--t distributions $\mathcal{ST}_p(\bb{\mu}, \bb{\Sigma}, \bb{\xi}, \nu)$, with $(\bb{\mu},\bb{\Sigma},\nu)=(\bb{0},{\rm{I}}_p, 12)$, and $\bb{\xi}=\theta \bb{1}_p$, with $\theta \in \{0,1,2,3,4,5,6\}$.  In this parameterization, $\theta=0$ corresponds to the null hypothesis and we are in the alternatives when $\theta>0$ (see the monograph \citealp{Azzalini2014} and the \textsf{R} package \texttt{sn} in \citealp{sn} for details).

\begin{figure}[!htb]
 \centering
 \includegraphics [scale=0.54]{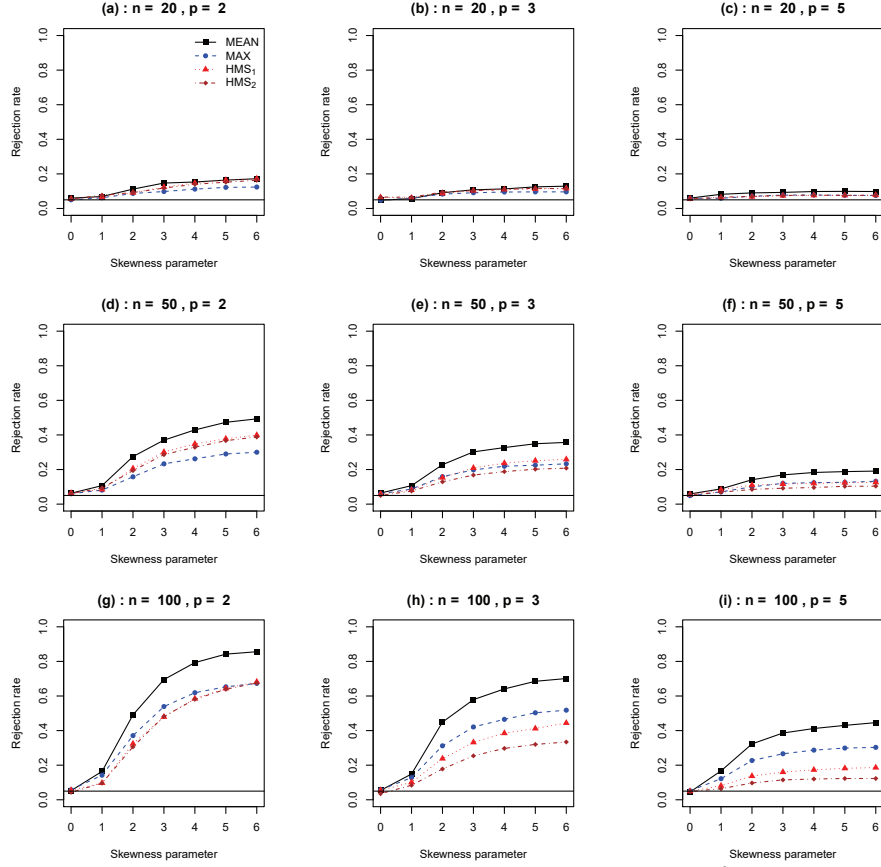}\par
 \vspace{-0.5cm}
 \caption{Simulation results for Example 3. Empirical rejection rates of $\overline {\widehat T}_{n,10}^{({\cal{N}})}$ (MEAN), $\widetilde {\widehat T}_{n,10}^{({\cal{N}})}$ (MAX), $HMS_1$, and $HMS_2$ against the skewness parameter $\theta$. The horizontal line corresponds to the $5\%$ significance level.}  \label{fig-mtt}
\end{figure}

The resulting empirical rejection rates for $\overline {\widehat T}_{n,10}^{({\cal{N}})}$, $\widetilde {\widehat T}_{n,10}^{({\cal{N}})}$, $HMS_1$, and $HMS_2$, are  plotted  against the skewness parameter $\theta$ in Figure \ref{fig-mtt}.
It can be seen clearly that the Type--I error rates ($\theta=0$) are controlled well for all scenarios, and the empirical rejection rates under alternatives ($\theta>0$) are in line with the fact that all the tests are consistent. In terms of comparative power, the proposed test $\overline {\widehat T}_{n,10}^{({\cal{N}})}$ (corresponding to the mean of the artificial test statistics) is clearly the most powerful for all scenarios. Moreover the other test $\widetilde {\widehat T}_{n,10}^{({\cal{N}})}$ corresponding to the maximum of the artificial test statistics also has higher empirical powers than the Wasserstein distance--based tests $HMS_1$ and $HMS_2$ when $n=100$.

\noindent
\textbf{Example 4: Tests for the Kotz--type distribution}

In this example, we apply the proposed tests to testing goodness--of--fit  for the
Kotz-type distribution $\mathcal{K}_p(\bb{\delta}, \bb{V}, N, s, r)$ with known parameters $N$, $s$, and $r$.
Here $\bb{\delta}$ and $\bb{V}$ denote the location vector and dispersion matrix respectively, and $N$, $s$, and $r$ are three scalar parameters that give model flexibility.
The Kotz--type distribution was introduced by \cite{Kotz1975} as a generalization of the multivariate normal distribution.
When $N=1$, $s=1$ and $r=1/2$, the distribution reduces to a multivariate normal distribution.
More detailed presentations can be found in \cite{Fang1990} and the review paper \cite{Nadarajah2003}.
We denote $\bb{X}\sim \mathcal{K}_p(\bb{\delta}, \bb{V}, N)$ for short if the $p$-dimensional random vector $\bb{X}$ follows the Kotz--type distribution with $s=1$ and $r=1/2$.
From Section 3.2.3 in \cite{Fang1990} we know that in this case $\mathbb{E}(\bb{X})=\bb{\delta}$ and $\mathbb{V}(\bb{X})=((2N+p-2)/p)\bb{V}$.
Then the parameters $\bb{\delta}$ and $\bb{V}$ can be estimated by the MEs $\widehat {\bb{\delta}}_n = \overline {\bb{X}}_n$ and $\widehat {\bb{V}}_n=(p/(2N+p-2)) \bb{S}_n$, respectively.

Now we apply the proposed tests to the goodness--of--fit test for $\mathcal{K}_p(\bb{0}, {\rm{I}}_p, 2)$ and generate observations from $\mathcal{K}_p(\bb{0}, {\rm{I}}_p, N)$ with $N\in \{1,1.5,2,2.5,3,3.5,4,4.5,5\}$.
Then $N=2$ corresponds to the null hypothesis, and we are in the alternatives for other values of $N$.
To simulate random observations from $\mathcal{K}_p(\bb{0}, {\rm{I}}_p, N)$, we use the stochastic representation given by \cite{Fang1990} in Section 3.2.2.
That is, $\bb{X}\stackrel{d}{=}R\bb{U}$ when $\bb{X}\sim\mathcal{K}_p(\bb{0}, {\rm{I}}_p, N)$, where $\stackrel{d}{=}$ stands for equality in distribution, $R^2$ follows the gamma distribution $Ga(N+p/2-1, 1/2)$, and $\bb{U}$ is uniformly distributed on the unit sphere and is independent of $R$.

\begin{figure}[!htb]
 \centering
 \includegraphics [scale=0.54]{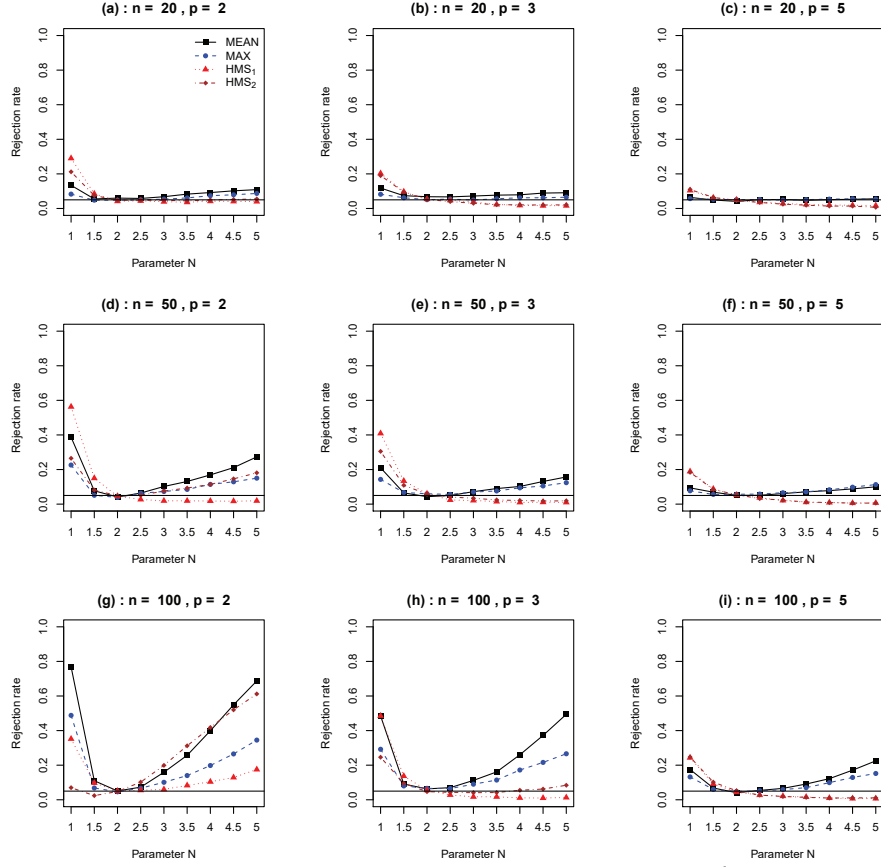}\par
 \vspace{-0.5cm}
 \caption{Simulation results for Example 4. Empirical rejection rates of $\overline {\widehat T}_{n,10}^{({\cal{N}})}$ (MEAN), $\widetilde {\widehat T}_{n,10}^{({\cal{N}})}$ (MAX), $HMS_1$, and $HMS_2$ against the parameter $N$. The horizontal line corresponds to the $5\%$ significance level.}  \label{fig-mkt}
\end{figure}

The empirical rejection rates corresponding to the tests $\overline {\widehat T}_{n,10}^{({\cal{N}})}$, $\widetilde {\widehat T}_{n,10}^{({\cal{N}})}$, $HMS_1$, and $HMS_2$, are  plotted  against the parameter $N$ in Figure \ref{fig-mkt}.
As we can see, the Type--I error rates for the null hypothesis ($N=2$) can be controlled well for all tests.
Meanwhile, as sample size $n$ increases and/or parameter $N$ goes away from $2$, the empirical powers of the proposed tests $\overline {\widehat T}_{n,10}^{({\cal{N}})}$ and $\widetilde {\widehat T}_{n,10}^{({\cal{N}})}$ increase with only an exception appearing when $p=5$ and $n=20$ (the highest dimension and smallest sample size).
However, the Wasserstein distance--based tests $HMS_1$ and $HMS_2$ perform quite differently from the proposed tests in view of the empirical power.
It can be seen clearly from Figure \ref{fig-mkt} that $HMS_1$ and $HMS_2$ only have non--trivial power against alternatives with $N=1$ and $N=1.5$ (lighter tail alternatives), while at the same time they either fail to detect and/or are clearly less powerful than the $\overline {\widehat T}_{n,10}^{({\cal{N}})}$ test for heavier tailed alternatives ($N>2$) in the great majority of cases. On the other hand the latter test, $\overline {\widehat T}_{n,10}^{({\cal{N}})}$, although having low power for $n=20$, its performance is enhanced with higher sample size, showing significant performance, particularly in lower dimensions, certainly when $N>2$ but also when $N<2$. We thus conclude that the proposed test $\overline {\widehat T}_{n,10}^{({\cal{N}})}$ is more robust  and therefore it should be preferred over
the $HMS_1$ and $HMS_2$ tests, particularly in the absence of any information regarding the type of departure from the null hypothesis (corresponding to varying values of the tail--parameter $N$).

\begin{rem} We also tried different weight functions leading to the test statistic in \eqref{ts1} with $\Psi(\cdot)$ other than the standard Gaussian CF. Specifically additional simulation experiments with spherical stable and Laplace weight functions were obtained. These results corroborate earlier findings, see \cite{Meintanis2005}, \cite{HM08} and \cite{JHMZ}, suggesting that the functional form of $w(t)$ is not all that crucial for test performance. These results are included in an accompanying online Supplement.     For more information related to the weight function we refer to the conclusions in Section \ref{sec8}.

\end{rem}

\subsection{A real--data set application} \label{realdata}
In this subsection, we apply the proposed tests to an ``Open-book Closed-book" data set (see \citealp{Mardia1979}, p. 3--4).
The data set contains 88 students' examination marks in Mechanics, Vectors, Algebra, Analysis, and Statistics.
\cite{Ducharme2020} recently performed a trivariate normality test on marks in Vectors, Algebra, and Statistics and concluded that the null hypothesis should be rejected due to the small $p$-value.
Here we test for 5-dimensional normality of all five marks.
As in the simulations, we consider the BHEP test and tests in \cite{Ducharme2020} and \cite{Hallin2021} for comparison.
Furthermore, to investigate the impact of randomness due to the artificial data in real--data analysis, we choose $m \in \{1,5,10,15,20,25,30\}$ as in the simulations too.
The empirical $p$-values based on 20000 resamples of $\overline {\widehat T}_{n,m}^{({\cal{N}})}$, $\widetilde {\widehat T}_{n,m}^{({\cal{N}})}$, $BHEP_1$, $DM$, $HMS_1$, and $HMS_2$ are presented in Table \ref{table-real}, and clearly suggest to reject the null hypothesis of normality at 5\% significance level. In this connection, it appears that the method suggested in order to deal with the randomness of artificial data in the new test and the approximation of critical points is feasible for real--data analysis.

\begin{table}[htbp]
  \centering
  \caption{The $p$-values of $\overline {\widehat T}_{n,m}^{({\cal{N}})}$ (MEAN), $\widetilde {\widehat T}_{n,m}^{({\cal{N}})}$ (MAX), $BHEP_1$, $DM$, $HMS_1$, and $HMS_2$ for the analyses of ``Open-book Closed-book" data set.} \label{table-real}
  \vspace{3mm}
  \renewcommand\arraystretch{1.5}
  \begin{tabular}{cccccccccccccccccccc}
  \cline{1-11}
    &\multicolumn{7}{c}{$\overline {\widehat T}_{n,m}^{({\cal{N}})}$ (MEAN)} & & & \\ \cline{2-8}
    Test &$m=1$ &$m=5$ &$m=10$ &$m=15$ &$m=20$ &$m=25$ &$m=30$ & &$BHEP_1$ &$DM$ \\ \cline{1-11}
    $p$-value &0.0015 &0.00005 &0 &0 &0 &0 &0 & &0 &0.0031 \\ \cline{1-11}
    &\multicolumn{7}{c}{$\widetilde {\widehat T}_{n,m}^{({\cal{N}})}$ (MAX)} & & & \\ \cline{2-8}
    Test &$m=1$ &$m=5$ &$m=10$ &$m=15$ &$m=20$ &$m=25$ &$m=30$ & &$HMS_1$ &$HMS_2$ \\ \cline{1-11}
    $p$-value &0.0005 &0.0049 &0.0095 &0.00315 &0.01075 &0.00395 &0.00185 & &0.0172 &0.00385 \\
   \cline{1-11}
\end{tabular}
\end{table}

\section{Conclusion} \label{sec8}
We propose goodness--of--fit tests that make use of the empirical characteristic function of the data at hand and its distance from its counterpart computed from data generated under the null hypothesis. This Monte Carlo sampling from the hypothesis under test allows for convenient explicit expression of the test statistic regardless of whether the null characteristic function is unknown or known but too complicated. The tests enjoy certain invariance properties, they are shown to be consistent and their limit distribution is investigated. Our asymptotic results show that compared to tests without resampling, the new tests involved extra variability resulting from this Monte Carlo sampling under the null. Thus in our simulation study the new tests are found somewhat less powerful than those tests. Nevertheless our tests exhibit a competitive performance against these standard tests, whenever such tests are available, and at the same time outperform other general--purpose but more complicated competitor tests in many sampling situations.

In this connection and as already mentioned the weight function $w(t)$ figuring in equation \eqref{tw} may in principle take arbitrary functional forms. Trivial conditions are that $w(t)$ should be non-negative and symmetric around zero. A further requirement, already mentioned in Section \ref{sec2} is computational convenience, meaning that it should render the test statistic in a closed--formula free of numerical integration. The most popular weight function, by far, has been the zero--mean normal density (or some variant thereof) and within this restricted context there is some work on how to choose an extra scale parameter that is related to the variance of this normal distribution. For this issue we refer to \cite{Hen97}, \cite{HJM},  \cite{Tenreiro09, Tenreiro19}, and \cite{AS15}. Now the more general issue of the proper choice of the specific functional form of the weight function has been investigated to some degree by \cite{LMR} and \cite{ALMM},  but the results are somewhat far from having an immediate impact on actual test implementation. We close on the note that although the emphasis herein is for elliptical families, the new tests may be applied to alternative testing situations (say with discrete distributions), and in more general settings, such as for instance in families of distributions over manifolds.

\section{Appendix} \label{sec9}
We provide proofs of the asymptotic results stated in Section \ref{sec5}. In doing so, the following notation will be used: ${\rm{o}}_{\mathbb P}(1)$ and $\overset{\mathcal{D}}{=}$
denote negligibility in probability and equality of distribution of random vectors, respectively; $X_n=O_{\mathbb P}(1)$ means that  the sequence of random variables $\{X_n\}$ is bounded in probability;
if ${\bb A}$ is a matrix, then $\|{\bb A}\|$ is the Frobenious norm (the Euclidean norm of the vector obtained by stacking the columns of ${\bb A}$).

\subsection{Preliminary results}
\begin{lem} \label{raiz.de.S}
Let $\bb{X}_1, \ldots, \bb{X}_n$, be i.i.d. copies of $\bb{X} \in \mathbb{R}^p$. Assume
that (i) $\bb{V}={\rm{diag}}(\lambda_1, \ldots, \lambda_p)$ with $\displaystyle \min_{1\leq j\leq p} \lambda_j>0$, (ii) $\widehat{\bb{V}}_n$  satisfies  \eqref{bahadur2}, and that (iii) $\widehat{\bb{V}}_n^{-1}$ exists. Then,
\begin{equation} \label{sin.tra}
 \sqrt{n}\left(\widehat{\bb{V}}_n^{-1/2}-{\bb{V}}^{-1/2}\right) = -\frac{1}{\sqrt{n}}\sum_{j=1}^n \bb{V}^{-1}\bb{L}_{\bb{\vartheta}}(\bb{X}_j)\bb{V}^{-1}\circ \bb{\Lambda}+{\rm{o}}_{\mathbb{P}}(1),
 \end{equation}
where $\bb{\Lambda}=(\bb{\Lambda}_{rv})$ with $\bb{\Lambda}_{rv} = \sqrt{\lambda_r}\sqrt{\lambda_v}/(\sqrt{\lambda_r}+\sqrt{\lambda_v})$, $1\leq r,v \leq p$. Moreover, if \eqref{inv} holds, then
\begin{equation} \label{con.tra}
\sqrt{n}\left(\widehat{\bb{V}}_n^{-1/2}-{\bb{V}}^{-1/2}\right)
= -\frac{1}{\sqrt{n}}\sum_{j=1}^n \bb{V}^{-1/2}{\bb L}_0({\bb Z}_j)\bb{V}^{-1/2}\circ {\bb \Lambda}+{\rm{o}}_{\mathbb{P}}(1),
 \end{equation}
 where ${\bb Z}_j=\bb{V}^{-1/2}(\bb{X}_j-\bb{\delta})$, $1 \leq j \leq n$.
\end{lem}
\noindent {\bf Proof} 
By a Taylor expansion,
\begin{equation}   \label{lemma1.aux1}
\sqrt{n}\left(\widehat{{\bb V}}_n^{-1}-{\bb V}^{-1}\right) = \frac{-1}{\sqrt{n}}\sum_{j=1}^n{\bb V}^{-1} {\bb L}_{\bb{\vartheta}}(\bb{X}_j) {\bb V}^{-1}+ {\rm{o}}_{\mathbb{P}}(1),
\end{equation}
and applying Theorem 1.1 in \cite{Del2018} we have,
 \[
 \sqrt{n}\left(\widehat{{\bb V}}_n^{-1/2}-{\bb V}^{-1/2}\right) = \sqrt{n}\int_0^{\infty}e^{-t{\bb V}^{-1/2}}\left(\widehat{{\bb V}}_n^{-1}-{\bb V}^{-1}\right) e^{-t{\bb V}^{-1/2}}{\rm{d}}t+{\bb R},
\]
where $\|{\bb R}\| \leq \sqrt{p} \displaystyle \min_j \lambda_j^{-3/2}\sqrt{n} \|\widehat{{\bb V}}_n^{-1}-{\bb V}^{-1}\|^2$. Now invoking \eqref{lemma1.aux1}, it follows that $\|{\bb R}\|= {\rm{o}}_{\mathbb{P}}(1)$ and that
\[
\sqrt{n}\int_0^{\infty}e^{-t{\bb V}^{-1/2}}\left(\widehat{{\bb V}}_n^{-1}-{\bb V}^{-1}\right) e^{-t{\bb V}^{-1/2}}{\rm{d}}t
=-\frac{1}{\sqrt{n}}\sum_{j=1}^n  {\bb V}^{-1}L_{\bb{\vartheta}}(\bb{X}_j){\bb V}^{-1}\circ {\bb \Lambda}+{\rm{o}}_{\mathbb{P}}(1),
\]
which proves \eqref{sin.tra}. To prove \eqref{con.tra}, we take into account that $\widehat{\bb V}_n^{-1}-{\bb V}^{-1}= 
{\bb V}^{-1/2}( \widehat{{\bb V}}_{n,{\bb Z}}^{-1}-{\rm{I}}_p){\bb V}^{-1/2}$, where $\widehat{{\bb V}}_{n,{\bb Z}}=\widehat{{\bb V}}_{n}({\bb Z}_1, \ldots, {\bb Z}_n)$, and then follow similar steps. $\Box$

\begin{lem} \label{los.Deltas}
Let $\bb{X}_1, \ldots, \bb{X}_n$ be i.i.d. copies of $\bb{X} \in \mathbb{R}^p$. Under the conditions
\begin{eqnarray*}
\widehat{\bb{\delta}}_n =\widehat{\bb{\delta}}_n (\bb{X}_1, \ldots, \bb{X}_n) & \overset{\mathbb{P}}{\to} & \bb{\delta}={\bb 0}, \\
\widehat{V}_n= \widehat{V}_n (\bb{X}_1, \ldots, \bb{X}_n)  & \overset{\mathbb{P}}{\to} & \bb{V},
\end{eqnarray*}
and that $ \widehat{\bb{V}}_n^{-1}$ exists, and letting $\bb{\Delta}_{n,j} = \widehat {\bb{X}}_j- \bb{V}^{-1/2} \bb{X}_j$, $1 \leq j \leq n$, we have,
\begin{enumerate} \itemsep=0pt
\item[{\rm{(a)}}] If  $\mathbb{E}(\|\bb{X}\|)<\infty$,  then $n^{-1}\sum_{j=1}^n \|\bb{\Delta}_{n,j}\| \overset{\mathbb{P}}{\to}  0$.
\item[{\rm{(b)}}] If $\mathbb{E}(\|\bb{X}\|^2)<\infty$, then $n^{-1}\sum_{j=1}^n \|\bb{\Delta}_{n,j}\|^2  \overset{\mathbb{P}}{\to} 0$.
\end{enumerate}
Moreover if  $\mathbb{E}(\|\bb{X}\|^2)<\infty$ and the conditions in Lemma \ref{raiz.de.S} hold, then
\begin{enumerate} \itemsep=0pt
\item[{\rm{(c)}}]  $n^{-1/2}\sum_{j=1}^n \|\bb{\Delta}_{n,j}\|^2 \overset{\mathbb{P}}{\to} 0$.   \end{enumerate}
\end{lem}

\noindent {\bf Proof} First of all we write
\begin{equation} \label{Deltanj}
\bb{\Delta}_{n,j} = \left( \widehat{\bb V}_n^{-1/2}- {\bb V}^{-1/2}\right)\bb{X}_j - \widehat{\bb V}_n^{-1/2}\widehat{\bb{\delta}}_n, \quad 1 \leq j \leq n.
\end{equation}
\noindent (a) We have that
\[
\frac{1}{n}\sum_{j=1}^n \|\bb{\Delta}_{n,j}\| \leq  \frac{1}{n}\sum_{j=1}^n \left\|  \left( \widehat{\bb V}_n^{-1/2}- {\bb V}^{-1/2}\right)\bb{X}_j \right\| + \left\| \widehat{\bb V}_n^{-1/2}\widehat{\bb{\delta}}_n \right\|.
\]
Since $\widehat{\bb V}_n  \overset{\mathbb{P}}{\to}  {\bb V}$ and $\widehat{\bb{\delta}}_n  \overset{\mathbb{P}}{\to} {\bb 0}$, it follows that $\|\widehat{\bb V}_n^{-1/2}\widehat{\bb{\delta}}_n \| \overset{\mathbb{P}}{\to} 0$. Also since $\|\widehat{\bb V}_n^{-1/2}- {\bb V}^{-1/2}\|  \overset{\mathbb{P}}{\to} 0$ and $n^{-1}\sum_{j=1}^n \| \bb{X}_j\|   \overset{\mathbb{P}}{\to} \mathbb{E}(\|{\bb X}\|)<\infty$, it follows that
 $$\frac{1}{n}\sum_{j=1}^n  \left\| \left(\widehat{\bb V}_n^{-1/2}- {\bb V}^{-1/2}\right)\bb{X}_j \right\| \leq \left\| \widehat{\bb V}_n^{-1/2}- {\bb V}^{-1/2} \right\| \frac{1}{n} \sum_{j=1}^n \| \bb{X}_j\| \overset{\mathbb{P}}{\to}  0,$$ which proves part (a).

\medskip
\noindent (b) The proof is similar to that of part (a), so we omit it.

\medskip

\noindent (c) Notice that
\[
 \|\bb{\Delta}_{n,j}\|^2 \leq  3\left\|  \left(\widehat{\bb V}_n^{-1/2}- {\bb V}^{-1/2}\right)\bb{X}_j\right\|^2 + 3\left\|\widehat{\bb V}_n^{-1/2}\widehat{\bb{\delta}}_n\right\|^2.
\]
Also from  Lemma \ref{raiz.de.S},  $\| \sqrt{n}(\widehat{\bb V}_n^{-1/2}- {\bb V}^{-1/2})\|={\rm{O}}_{\mathbb{P}}(1)$, and by taking into account the WLLN which entails $n^{-1}\sum_{j=1}^n\|\bb{X}_j\|^2  \overset{\mathbb{P}}{\to}  \mathbb{E}(\|{\bb X}\|^2)<\infty$, we have
\[
\frac{1}{\sqrt{n}}\sum_{j=1}^n \left\| \left(\widehat{\bb V}_n^{-1/2}- {\bb V}^{-1/2}\right)\bb{X}_j \right\|^2 \leq \frac{1}{\sqrt{n}} \left\| \sqrt{n} \left(
\widehat{\bb V}_n^{-1/2} - {\bb V}^{-1/2}\right)\right\|^2 \frac{1}{n}\sum_{j=1}^n \|\bb{X}_j\|^2 ={\rm{o}}_{\mathbb{P}}(1).
\]
Moreover the continuous mapping theorem implies that $\|\widehat{\bb V}_n^{-1/2}\|  \overset{\mathbb{P}}{\to}  \|{\bb V}^{-1/2}\|<\infty$, and by taking into account the CLT and the continuous mapping theorem, we obtain  $\|\sqrt{n}\widehat{\bb{\delta}}_n\|={\rm{O}}_{\mathbb{P}}(1)$. Consequently we have
\[
\sqrt{n} \left\|\widehat{\bb V}_n^{-1/2}\widehat{\bb{\delta}}_n\right\|^2 \leq \frac{1}{\sqrt{n}} \left\|\widehat{\bb V}_n^{-1/2}\right\|^2 \left\|\sqrt{n}\hat{\bb{\delta}}_n\right\|^2 ={\rm{o}}_{\mathbb{P}}(1),
\]
which together with the previous two equations  proves part (c). $\Box$

Proposition 3 of \cite{Dorr2021} (see also Proposition 1 in \citealp{Ebner2020}) is a special case of Lemma \ref{los.Deltas} for the case where $\widehat{\bb{\delta}}_n$ and $\widehat{\bb V}_n$ are replaced by the sample mean $\overline {\bb X}_n$ and the sample variance ${\bb S}_n$, respectively, and
 $\mathbb{E}({\bb X}{\bb X}^\top)={\rm{I}}_p$.

\begin{lem} \label{lemma3}
Let ${\bb X}_1, \ldots, {\bb X}_n$, be i.i.d. copies of ${\bb X} \in \mathbb{R}^p$. Assume that the assumptions
in Lemma \ref{los.Deltas} (c) hold, and that  \eqref{equiv}, \eqref{inv}, \eqref{bahadur2},   and $\int \|{\bb t}\|^4w({\bb  t}){\rm{d}}{\bb t}<\infty$ also hold. Let ${\bb Y}_j={\bb V}^{-1/2}\bb{X}_j$, $1 \leq j \leq n$, and ${\bb Y}={\bb V}^{-1/2}{\bb X}$. Let ${\bb \Lambda}$ be as defined in the statement of Lemma \ref{raiz.de.S}. Then,
\begin{enumerate} \itemsep=0pt
\item[{\rm{(a)}}] $\displaystyle
 \frac{1}{\sqrt{n}}\sum_{j=1}^n\left\{\cos({\bb  t}^\top \widehat {\bb X}_j)-\cos({\bb  t}^\top {\bb Y}_j)\right\}={\bb  t}^\top\frac{1}{\sqrt{n}}\sum_{j=1}^n C_{{\bb Y}_j}({\bb  t})+r_{c,n}({\bb  t})$, with 
\[C_{\bb Y}({\bb  t})={\rm Im}(\varphi_{\bb Y}({\bb  t})) \bb{\ell}_0({\bb Y})-{\bb L}_0({\bb Y})\circ {\bb V}^{-1/2}{\bb \Lambda} \frac{\partial}{\partial {\bb  t}} {\rm Re}(\varphi_{\bb Y}({\bb  t})),
\] and  $\|r_{c,n}\|_w={\rm{o}}_{\mathbb{P}}(1)$.
\item[{\rm{(b)}}] $\displaystyle
 \frac{1}{\sqrt{n}}\sum_{j=1}^n\left\{\sin({\bb  t}^\top \widehat {{\bb X}}_j)-\sin({\bb  t}^\top {\bb Y}_j)\right\}={\bb t}^\top\frac{1}{\sqrt{n}}\sum_{j=1}^n S_{{\bb Y}_j}({\bb  t})+r_{s,n}({\bb  t})$, with 
\[
S_{\bb Y}({\bb  t})=-{\rm Re}(\varphi_{\bb Y}({\bb  t})) \bb{\ell}_0({\bb Y})-{\bb L}_0({\bb Y})\circ {\bb V}^{-1/2}{\bb \Lambda}\frac{\partial}{\partial {\bb  t}}  {\rm Im}(\varphi_{\bb Y}({\bb  t})),
\]
and $\|r_{s,n}\|_w={\rm{o}}_{\mathbb{P}}(1)$.
\end{enumerate}
\end{lem}
\noindent {\bf Proof} (a) By Taylor expansion
\[
\frac{1}{\sqrt{n}}\sum_{j=1}^n\left\{\cos({\bb  t}^\top \widehat {{\bb X}}_j)-\cos({\bb  t}^\top {\bb Y}_j)\right\}={\bb  t}^\top\frac{-1}{\sqrt{n}}\sum_{j=1}^n \sin({\bb  t}^\top {\bb Y}_j)
\bb{\Delta}_{n,j}+r_{1,n}({\bb  t}),
\]
with
\[
\|r_{1,n}\|_w \leq \left( \int \|{\bb  t}\|^4w({\bb  t}){\rm d}{\bb  t} \right)^{1/2} \frac{1}{\sqrt{n}}\sum_{j=1}^n \|\bb{\Delta}_{n,j}\|^2.
\]
From Lemma  \ref{los.Deltas} (c), it follows that $\|r_{1,n}\|_w={\rm{o}}_{\mathbb{P}}(1)$.  Recall expression \eqref{Deltanj}. Since $n^{-1}\sum_{j=1}^n \sin({\bb  t}^\top {\bb Y}_j) \overset{\mathbb{P}}{\to}  {\rm Im}(\varphi_{\bb Y}({\bb  t}))$,
$ \widehat{{\bb V}}_n^{-1/2} \overset{\mathbb{P}}{\to} {\bb V}^{-1/2}$, \eqref{equiv} and \eqref{bahadur1},  we get that
\[
{\bb  t}^\top \frac{1}{\sqrt{n}}\sum_{j=1}^n \sin({\bb  t}^\top {\bb Y}_j)  \widehat{{\bb V}}_n^{-1/2} \widehat{\bb{\delta}}_n = {\rm Im}(\varphi_{\bb Y}({\bb  t})) {\bb  t}^\top \frac{1}{\sqrt{n}}\sum_{j=1}^n \bb{\ell}_0({\bb Y}_j)+r_{2,n}({\bb  t}),
\]
with $\|r_{2,n}\|_w={\rm{o}}_{\mathbb{P}}(1)$. Using \eqref{inv},
Lemma  \ref{raiz.de.S} and taking into account that $-n^{-1}\sum_{j=1}^n \sin({\bb  t}^\top {\bb Y}_j)\bb{X}_j  \overset{\mathbb{P}}{\to} {\bb V}^{1/2} \partial{\rm Re}(\varphi_{\bb Y}({\bb  t})) / \partial {\bb  t}$,
we get that
\[
{\bb  t}^\top \frac{1}{\sqrt{n}}\sum_{j=1}^n \sin({\bb  t}^\top {\bb Y}_j) ( \widehat{{\bb V}}_n^{-1/2}  - {\bb V}^{-1/2}) \bb{X}_j=
{\bb  t}^\top \frac{1}{\sqrt{n}}\sum_{j=1}^n {\bb V}^{-1/2}{\bb L}_0({\bb Y}_j){\bb V}^{-1/2}\circ {\bb \Lambda} {\bb V}^{1/2}\frac{\partial}{\partial {\bb  t}} {\rm Re}(\varphi_{\bb Y}({\bb  t}))+r_{3,n}({\bb  t})
\]
with $\|r_{3,n}\|_w={\rm{o}}_{\mathbb{P}}(1)$.  Now, taking into account that ${\bb V}^{-1/2}{\bb L}_0({\bb Y}){\bb V}^{-1/2}\circ {\bb \Lambda} {\bb V}^{1/2}={\bb L}_0({\bb Y}) \circ {\bb V}^{-1/2}{\bb \Lambda}$, part (a) has been proven.

\medskip

\noindent The proof of part (b) is parallel, so we omit it. $\Box$

\begin{rem} \label{nota1} In the context of Lemma \ref{lemma3},
if ${\bb Y}={\bb V}^{-1/2}{\bb X}$ has a spherically symmetric distribution with CF $\varphi_0({\bb t})=\Psi_0(\|{\bb t}\|^2)$, for some function $\Psi_0(\cdot)$ of a scalar variable, then the expressions of $C_{\bb Y}({\bb t})$ and $S_{\bb Y}({\bb t})$ may be written as
\begin{eqnarray*}
C_{\bb Y}({\bb t}) & = & -2\Psi_0'(\|{\bb t}\|^2){\bb L}_0({\bb Y})\circ {\bb V}^{-1/2} {\bb \Lambda } {\bb t},\\ 
S_{\bb Y}({\bb t}) & = & -\Psi_0(\|{\bb t}\|^2)\bb{\ell}_0({\bb Y}),
\end{eqnarray*}
where $\Psi_0'(x)={\rm{d}} \Psi_0(x)/{\rm{d}}x$.
\end{rem}

\subsection{Proofs of main results}

\noindent {\bf Proof of Theorem \ref{limit}}
From \eqref{sym}, it follows that
\[
\kappa=\|{\rm Re}(\varphi_{{\bb Y}}) + {\rm Im}(\varphi_{{\bb Y}}) - {\rm Re}(\varphi_0) - {\rm Im}(\varphi_0)\|_w^2.
\] 
Let $\varphi_{1,n}$ denote the empirical CF of ${\bb V}^{-1/2}{\bb X}_1, \ldots, {\bb V}^{-1/2}{\bb X}_n$.
First, we will see that
\begin{equation} \label{aux1}
\|{\rm Re}(\varphi_n) + {\rm Im}(\varphi_n) - {\rm Re}(\varphi_{1,n}) - {\rm Im}(\varphi_{1,n})\|_w \overset{\mathbb{P}}{\to} 0.
\end{equation}
Since
$|\cos(x)-\cos(y)| \leq |x-y|$ and
$|\sin(x)-\sin(y)| \leq |x-y|
$, $\forall x, y \in \mathbb{R}$,
it follows that
$$
|{\rm Re}(\varphi_n({\bb t})) - {\rm Re}(\varphi_{1,n}({\bb t}))| \leq \|{\bb t}\| \frac{1}{n}\sum_{j=1}^n \|\bb{\Delta}_{n,j}\|, \quad
|{\rm Im}(\varphi_n({\bb t})) - {\rm Im}(\varphi_{1,n}({\bb t}))| \leq \|{\bb t}\| \frac{1}{n}\sum_{j=1}^n \|\bb{\Delta}_{n,j}\|,
$$
with $\bb{\Delta}_{n,j}$ as defined in the statement of Lemma  \ref{los.Deltas}. Applying Lemma \ref{los.Deltas} (a), it follows that
 \eqref{aux1} holds. Now, the WLLN in Hilbert spaces and the continuous mapping theorem imply that
\begin{equation} \label{aux2}
\| {\rm Re}(\varphi_{1,n}) + {\rm Im}(\varphi_{1,n}) - {\rm Re}(\varphi_{0}) - {\rm Im}(\varphi_{0})\|_w \overset{\mathbb{P}}{\to}  \kappa.
\end{equation}
Then the result follows from  \eqref{aux1},  \eqref{aux2} and \eqref{equality}. $\Box$

\medskip

\noindent {\bf Proof of Theorem \ref{asympt.null.distrib}} Recall that $\widehat T^{(\Psi)}_{2,n}$ can be expressed as in \eqref{T2}. Now, using Lemma \ref{lemma3} and Remark \ref{nota1}, we can write
\[
G_n({\bb t})\overset{\mathcal{D}}{=} \frac{1}{\sqrt{n}}\sum_{j=1}^n \left\{W_1({\bb Y}_{0,j},V, \Psi_0; {\bb t}) - W_2({\bb X}_{0,j}, \Psi_0; {\bb t}) \right\} + r_n({\bb t}),
\]
 where $\|r_n\|_w={\rm{o}}_{\mathbb{P}}(1)$ and $\{{\bb Y}_{0,1}, \ldots, {\bb Y}_{0,n}, {\bb X}_{0,1}, \ldots, {\bb X}_{0,n}\}$ is a set of i.i.d. observations on ${\bb X}_0$. Now the result follows from the CLT in separable Hilbert spaces, the continuous mapping theorem and \eqref{equality}.  $\Box$

\medskip

\noindent {\bf Proof of Proposition \ref{la.cova}}  Standard calculations show that
\begin{equation} \label{p.1}
\begin{array}{c}
\mathbb{E}\Big[\left\{\cos({\bb t}^\top {\bb X}_0)-\varphi_0({\bb t}^\top {\bb t})+\sin({\bb t}^\top {\bb X}_0)\right\} \left\{\cos({\bb s}^\top {\bb X}_0)-\varphi_0({\bb s}^\top {\bb s})+\sin({\bb s}^\top {\bb X}_0)\right\} \Big]\\
= \varphi_0({\bb t}-{\bb s})-\varphi_0({\bb t})\varphi_0({\bb s}). \end{array}
\end{equation}
Since ${\bb L}_0({\bb X}_0)= {\bb L}_0(-{\bb X}_0)$, it follows that  $\mathbb{E}\left\{ {\bb L}_0({\bb X}_0) \sin(s^\top {\bb X}_0) \right\}={\bb 0}$, and thus
\begin{equation} \label{p.2}
\mathbb{E}\left\{ {\bb L}_0({\bb X}_0)\circ {\bb V}^{-1/2} {\bb \Lambda} \sin({\bb s}^\top {\bb X}_0) \right\}={\bb 0},
\end{equation}
where ${\bb \Lambda}$  is as defined in the statement of Lemma \ref{raiz.de.S}.
Likewise, since  $\bb{\ell}_0(-{\bb X}_0)= -\bb{\ell}_0({\bb X}_0)$ and  $\mathbb{E}\left\{ \bb{\ell}_0({\bb X}_0)\right\}={\bb 0}$, we have that
\begin{equation} \label{p.3}
\mathbb{E}\left[\bb{\ell}_0({\bb Y}) \left\{ \cos({\bb t}^\top {\bb X})-\varphi_0({\bb t}^\top {\bb t}) \right\}\right]={\bb 0}.
\end{equation}
Also the conditions ${\bb L}_0({\bb X}_0)= {\bb L}_0(-{\bb X}_0)$ and $\bb{\ell}_0(-{\bb X}_0)= -\bb{\ell}_0({\bb X}_0)$, entail
\begin{equation} \label{p.4}
\mathbb{E}\left\{ {\bb L}_0({\bb Y})\circ {\bb V}^{-1/2} {\bb \Lambda}  {\bb t} {\bb s}^\top \bb{\ell}_0({\bb Y}) \right\}={\bb 0}.
\end{equation}
From \eqref{condition} and taking into account that $\mathbb{E}\{{\bb L}_0({\bb X}_0)\}={\bb 0}$, we get
\begin{equation} \label{p.5}
\mathbb{E}\left[ {\bb t}^\top {\bb L}_0({\bb X}_0)\circ {\bb V}^{-1/2} {\bb \Lambda}  {\bb t} \left\{ \cos({\bb s}^\top {\bb X}_0)-\varphi_0({\bb s}^\top {\bb s}) \right\} \right]=
{\bb t}^\top \left\{\alpha({\bb s}){\rm I}_p+\beta({\bb s}){\bb s}{\bb s}^\top \right\} \circ {\bb V}^{-1/2} {\bb \Lambda}  {\bb t}.
\end{equation}
Let ${\bb t}^\top=(t_1, \ldots, t_p)$, ${\bb s}^\top=(s_1, \ldots, s_p)$, and ${\bb A}={\bb V}^{-1/2} {\bf \Lambda}=(a_{jk})$. By noticing that  $a_{jj}=1/2$ and $a_{jk}=\sqrt{\lambda_j}/(\sqrt{\lambda_j}+\sqrt{\lambda_k})$, $1\leq j \neq k \leq p$, it follows that
\begin{eqnarray}
{\bb t}^\top {\rm I}_p \circ {\bb V}^{-1/2} {\bb \Lambda}  {\bb t} & = & \frac{1}{2} {\bb t}^\top  {\bb t}, \label{p.6}\\
  {\bb t}^\top  {\bb s} {\bb s}^\top \circ  {\bb V}^{-1/2} {\bb \Lambda }  {\bb t} & = & \sum_{j,k=1}^p t_jt_ks_js_ka_{jk}
  =   \sum_{j,k=1}^p t_j^2s_j^2 a_{jj}+\sum_{j<k} t_jt_ks_js_k\left\{a_{jk}+a_{kj}\right\} \nonumber \\
 & = & \frac{1}{2}( {\bb t}^\top  {\bb s})^2. \label{p.7}
\end{eqnarray}
From the assumptions made, ${\bb L}_0({\bb X}_0)$ is a rotationally invariant matrix, that is, ${\bb L}_0({\bb X}_0)\overset{\mathcal{D}}{=}{\bb U} {\bb L}_0({\bb X}_0){\bb U}^\top$, $\forall {\bb U} \in \mathcal{O}_p$. Then, from Proposition 13.2 in \cite{Bilodeau1999}, all off-diagonal elements of ${\bb L}_0({\bb X}_0)$ are uncorrelated with each other and uncorrelated with the diagonal elements, i.e.
\begin{eqnarray*}
\mathbb{E}\left\{ {\bb L}_0({\bb X}_0)_{jj}^2 \right\} & = & 2\sigma_1+\sigma_2,\\
\mathbb{E}\left\{ {\bb L}_0({\bb X}_0)_{jj}{\bb L}_0({\bb X}_0)_{kk} \right\} & = & \sigma_2,\\
\mathbb{E}\left\{ {\bb L}_0({\bb X}_0)_{jk}^2 \right\} & = & \sigma_1,
\end{eqnarray*}
$1\leq j \neq k \leq p$.
Therefore,
\begin{equation}\label{p.8}
\begin{array}{l}
 \mathbb{E}\left[ {\bb t}^\top {\bb L}_0({\bb X}_0)\circ {\bb V}^{-1/2} {\bb \Lambda}  {\bb t} {\bb s}^\top {\bb L}_0({\bb X}_0)\circ {\bb V}^{-1/2} {\bb \Lambda}  {\bb s} \right]\\
\quad =\sum_{j,k=1}^p\sum_{u,v=1}^p t_jt_ks_us_va_{jk}a_{uv}\mathbb{E}\left\{ {\bb L}_0({\bb X}_0)_{jk}{\bb L}_0({\bb X}_0)_{uv} \right\}  \\
\quad =(2\sigma_1+\sigma_2)\sum_{j=1}^p t_j^2s_j^2a_{jj}^2+\sigma_1 \sum_{j \neq k}t_jt_ks_js_k(a_{jk}^2+a_{jk}a_{kj})+\sigma_2 \sum_{j \neq k}t_j^2s_k^2a_{jj}a_{kk}\\
\quad
=\frac{\sigma_1}{2} ({\bb t}^\top {\bb s})^2+\frac{\sigma_2}{4} {\bb t}^\top {\bb t} {\bb s}^\top {\bb s}.
\end{array}
\end{equation}
The result follows from \eqref{p.1}--\eqref{p.8}. $\Box$

\medskip

\noindent {\bf Proof of Theorem \ref{power}}  The result follows from \eqref{equality}, Lemma  \ref{lemma3} and  Theorem 1 in \cite{Baringhaus2017}. $\Box$

\section*{Acknowledgements}
The authors thank the Associate Editor and two anonymous reviewers for their constructive comments and suggestions which led to improvement of the presentation.
This research was supported by the National Natural Science Foundation of China [Grant No.12101055, No.12131006].
M.D. Jim\'enez-Gamero has been partially supported by grants PID2020-118101GB-I00, Ministerio de Ciencia e Innovaci\'on (MCIN/ AEI /10.13039/501100011033) and  P18-FR-2269 (Junta de Andaluc\'ia).

\bibliography{gof}

\begin{thebibliography}{56}
\expandafter\ifx\csname natexlab\endcsname\relax\def\natexlab#1{#1}\fi
\expandafter\ifx\csname url\endcsname\relax
  \def\url#1{\texttt{#1}}\fi
\expandafter\ifx\csname urlprefix\endcsname\relax\def\urlprefix{URL }\fi

\bibitem[{Alba-Fern\'{a}ndez et~al.(2017)Alba-Fern\'{a}ndez, Batsidis,
  Jim\'{e}nez-Gamero, and Jodr\'{a}}]{Alba2017}
Alba-Fern\'{a}ndez, M.~V., Batsidis, A., Jim\'{e}nez-Gamero, M.~D., Jodr\'{a},
  P., 2017. A class of tests for the two-sample problem for count data. J.
  Comput. Appl. Math. 318, 220--229.

\bibitem[{Alba-Fern\'{a}ndez et~al.(2008)Alba-Fern\'{a}ndez,
  Jim\'{e}nez-Gamero, and Mu\~{n}oz Garc\'{\i}a}]{Alba2008}
Alba-Fern\'{a}ndez, V., Jim\'{e}nez-Gamero, M.~D., Mu\~{n}oz Garc\'{\i}a, J.,
  2008. A test for the two-sample problem based on empirical characteristic
  functions. Comput. Statist. Data Anal. 52~(7), 3730--3748.

\bibitem[{Albert et~al.(2022)Albert, Laurent, Marrel, and Meynaoui}]{ALMM}
Albert, M., Laurent, B., Marrel, A., Meynaoui, A., 2022. Adaptive test of
  independence based on hsic measures. Ann. Statist. 50, 858--879.

\bibitem[{Allison and Santana(2015)}]{AS15}
Allison, J., Santana, L., 2015. On a data--dependent choice of the tuning
  parameter appearing in certain goodness--of--fit tests. J. Statist. Comput.
  Simul. 85, 3276--3288.

\bibitem[{Azzalini(2014)}]{Azzalini2014}
Azzalini, A., 2014. The skew-normal and related families. Vol.~3 of Institute
  of Mathematical Statistics (IMS) Monographs. Cambridge University Press,
  Cambridge, with the collaboration of Antonella Capitanio.

\bibitem[{Azzalini(2022)}]{sn}
Azzalini, A., 2022. The {R} package \texttt{sn}: The skew-normal and related
  distributions such as the skew-$t$ and the {SUN} (version 2.0.2).
  Universit\`a degli Studi di Padova, Italia, home page:
  \url{http://azzalini.stat.unipd.it/SN/}.

\bibitem[{Baringhaus et~al.(2017)Baringhaus, Ebner, and Henze}]{Baringhaus2017}
Baringhaus, L., Ebner, B., Henze, N., 2017. The limit distribution of weighted
  {$L^2$}-goodness-of-fit statistics under fixed alternatives, with
  applications. Ann. Inst. Statist. Math. 69~(5), 969--995.

\bibitem[{Baringhaus and Franz(2004)}]{BF04}
Baringhaus, L., Franz, C., 2004. On a new multivariate two--sample test. J.
  Multivar. Anal. 88, 190--206.

\bibitem[{Baringhaus and Henze(1988)}]{Baringhaus1988}
Baringhaus, L., Henze, N., 1988. A consistent test for multivariate normality
  based on the empirical characteristic function. Metrika 35~(6), 339--348.

\bibitem[{Bilodeau and Brenner(1999)}]{Bilodeau1999}
Bilodeau, M., Brenner, D., 1999. Theory of multivariate statistics. Springer
  Texts in Statistics. Springer-Verlag, New York.

\bibitem[{Chen et~al.(2019)Chen, Meintanis, and Zhu}]{Chen2019}
Chen, F., Meintanis, S.~G., Zhu, L., 2019. On some characterizations and
  multidimensional criteria for testing homogeneity, symmetry and independence.
  J. Multivariate Anal. 173, 125--144.

\bibitem[{Cuesta-Albertos et~al.(2006)Cuesta-Albertos, Fraiman, and
  Ransford}]{Cuesta2006}
Cuesta-Albertos, J.~A., Fraiman, R., Ransford, T., 2006. Random projections and
  goodness-of-fit tests in infinite-dimensional spaces. Bull. Braz. Math. Soc.
  (N.S.) 37~(4), 477--501.
\newline\urlprefix\url{https://doi.org/10.1007/s00574-006-0023-0}

\bibitem[{Del~Moral and Niclas(2018)}]{Del2018}
Del~Moral, P., Niclas, A., 2018. A {T}aylor expansion of the square root matrix
  function. J. Math. Anal. Appl. 465~(1), 259--266.

\bibitem[{D\"{o}rr et~al.(2021)D\"{o}rr, Ebner, and Henze}]{Dorr2021}
D\"{o}rr, P., Ebner, B., Henze, N., 2021. Testing multivariate normality by
  zeros of the harmonic oscillator in characteristic function spaces. Scand. J.
  Stat. 48~(2), 456--501.

\bibitem[{Ducharme and Lafaye~de Micheaux(2019)}]{ECGofTestDx}
Ducharme, G.~R., Lafaye~de Micheaux, P., 2019. ECGofTestDx: A Goodness-of-fit
  Test for Elliptical Distributions with Diagnostic Capabilities. R package
  version 0.4.

\bibitem[{Ducharme and Lafaye~de Micheaux(2020)}]{Ducharme2020}
Ducharme, G.~R., Lafaye~de Micheaux, P., 2020. A goodness-of-fit test for
  elliptical distributions with diagnostic capabilities. J. Multivariate Anal.
  178, 104602, 13.

\bibitem[{Ebner and Henze(2020)}]{Ebner2020}
Ebner, B., Henze, N., 2020. Tests for multivariate normality---a critical
  review with emphasis on weighted {$L^2$}-statistics. TEST 29~(4), 845--892.

\bibitem[{Epps and Singleton(1986)}]{ES86}
Epps, T., Singleton, K.~J., 1986. An omnibus test for the two-sample problem
  using the empirical characteristic function. J. Statist. Computat. Simul. 26,
  177--203.

\bibitem[{Epps and Pulley(1983)}]{Epps1983}
Epps, T.~W., Pulley, L.~B., 1983. A test for normality based on the empirical
  characteristic function. Biometrika 70~(3), 723--726.

\bibitem[{Fan(1997)}]{Fan1997}
Fan, Y., 1997. Goodness-of-fit tests for a multivariate distribution by the
  empirical characteristic function. J. Multivariate Anal. 62~(1), 36--63.

\bibitem[{Fang et~al.(1990)Fang, Kotz, and Ng}]{Fang1990}
Fang, K.~T., Kotz, S., Ng, K.~W., 1990. Symmetric multivariate and related
  distributions. Vol.~36 of Monographs on Statistics and Applied Probability.
  Chapman and Hall, Ltd., London.

\bibitem[{Fragiadakis and Meintanis(2011)}]{Fragiadakis2011}
Fragiadakis, K., Meintanis, S.~G., 2011. Goodness-of-fit tests for multivariate
  {L}aplace distributions. Math. Comput. Modelling 53~(5-6), 769--779.

\bibitem[{Gretton et~al.(2012)Gretton, Borgwardt, Rasch, and
  Sch\"olkopf}]{GBRSS}
Gretton, A., Borgwardt, K., Rasch, M., Sch\"olkopf, B.~Smola, A., 2012. A
  kernel two-sample test. J. Mach. Learn. Resear. 13, 723--773.

\bibitem[{Hallin et~al.(2021)Hallin, Mordant, and Segers}]{Hallin2021}
Hallin, M., Mordant, G., Segers, J., 2021. Multivariate goodness-of-fit tests
  based on {W}asserstein distance. Electron. J. Stat. 15~(1), 1328--1371.

\bibitem[{Henze(1997)}]{Hen97}
Henze, N., 1997. Extreme smoothing and testing for multivariate normality.
  Statist. Probab. Lett. 35, 203--213.

\bibitem[{Henze(2002)}]{Henze2002}
Henze, N., 2002. Invariant tests for multivariate normality: a critical review.
  Statist. Papers 43~(4), 467--506.

\bibitem[{Henze et~al.(2019)Henze, Jim\'enez-Gamero, and Meintanis}]{HJM}
Henze, N., Jim\'enez-Gamero, M., Meintanis, S., 2019. Characterizations of
  multinormality and corresponding tests of fit, including for garch models.
  Econometr. Theor. 35, 510--546.

\bibitem[{Henze et~al.(2005)Henze, Klar, and Zhu}]{Henze2005}
Henze, N., Klar, B., Zhu, L.~X., 2005. Checking the adequacy of the
  multivariate semiparametric location shift model. J. Multivariate Anal.
  93~(2), 238--256.

\bibitem[{Henze and Wagner(1997)}]{Henze1997}
Henze, N., Wagner, T., 1997. A new approach to the {BHEP} tests for
  multivariate normality. J. Multivariate Anal. 62~(1), 1--23.

\bibitem[{Henze and Zirkler(1990)}]{HZ90}
Henze, N., Zirkler, B., 1990. A class of invariant consistent tests for
  multivariate normality. Commun. Statist.--Theor. Meth. 19, 3595--3617.

\bibitem[{Huber-Carol et~al.(2002)Huber-Carol, Balakrishnan, Nikulin, and
  Mesbah}]{HBNM}
Huber-Carol, C., Balakrishnan, N., Nikulin, M., Mesbah, M., 2002.
  Goodness--of--Fit Tests and Model Validity. Birkh\"auser, Boston.

\bibitem[{Hu\v{s}kov\'a and Meintanis(2008)}]{HM08}
Hu\v{s}kov\'a, M., Meintanis, S., 2008. Tests for the multivariate k-sample
  problem based on the empirical characteristic function. J. NonParam. Statist.
  20, 263--277.

\bibitem[{Jiang et~al.(2019)Jiang, Hu\v{s}kov\'a, Meintanis, and Zhu}]{JHMZ}
Jiang, Q., Hu\v{s}kov\'a, M., Meintanis, S., Zhu, L., 2019. Asymptotics,
  finite-sample comparisons and applications for two-sample tests with
  functional data. J. Multivar. Anal. 170, 202--220.

\bibitem[{Jim\'{e}nez-Gamero et~al.(2017)Jim\'{e}nez-Gamero,
  Alba-Fern\'{a}ndez, Jodr\'{a}, and Barranco-Chamorro}]{JG2017}
Jim\'{e}nez-Gamero, M., Alba-Fern\'{a}ndez, M.~V., Jodr\'{a}, P.,
  Barranco-Chamorro, I., 2017. Fast tests for the two-sample problem based on
  the empirical characteristic function. Math. Comput. Simulation 137,
  390--410.
\newline\urlprefix\url{https://doi.org/10.1016/j.matcom.2016.09.007}

\bibitem[{Jim\'{e}nez-Gamero et~al.(2009)Jim\'{e}nez-Gamero,
  Alba-Fern\'{a}ndez, Mu\~{n}oz Garc\'{\i}a, and Chalco-Cano}]{JG2009}
Jim\'{e}nez-Gamero, M.~D., Alba-Fern\'{a}ndez, V., Mu\~{n}oz Garc\'{\i}a, J.,
  Chalco-Cano, Y., 2009. Goodness-of-fit tests based on empirical
  characteristic functions. Comput. Statist. Data Anal. 53~(12), 3957--3971.
\newline\urlprefix\url{https://doi.org/10.1016/j.csda.2009.06.001}

\bibitem[{Kelker(1970)}]{Kelker70}
Kelker, D., 1970. Distribution theory of spherical distributions and a
  location--scale parameter generalization. Sankhy$\bar {\rm{a}}$ A 32,
  419--430.

\bibitem[{Kotz(1975)}]{Kotz1975}
Kotz, S., 1975. Multivariate Distributions at a Cross Road. In: Patil G.P.,
  Kotz S., Ord J.K. (eds) A Modern Course on Statistical Distributions in
  Scientific Work. NATO Advanced Study Institutes Series (Series C ¡ª
  Mathematical and Physical Sciences). Vol.~17. Springer, Dordrecht.

\bibitem[{Kotz and Nadarajah(2004)}]{Kotz2004}
Kotz, S., Nadarajah, S., 2004. Multivariate {$t$} distributions and their
  applications. Cambridge University Press, Cambridge.

\bibitem[{Koutrouvelis and Kellermeier(1981)}]{KK81}
Koutrouvelis, I., Kellermeier, J., 1981. A goodness--of--fit test based on the
  empirical characteristic function when parameters must be estimated. J. Roy.
  Statist. Soc. B 43, 173--176.

\bibitem[{Koutrouvelis and Meintanis(1999)}]{MK99}
Koutrouvelis, I., Meintanis, S., 1999. Testing for stability based on the
  empirical characteristic function with applications to financial data. J.
  Statist. Computat. Simul. 64, 275--300.

\bibitem[{Kozubowski et~al.(2013)Kozubowski, Podg\'{o}rski, and
  Rychlik}]{Kozubowski2013}
Kozubowski, T.~J., Podg\'{o}rski, K., Rychlik, I., 2013. Multivariate
  generalized {L}aplace distribution and related random fields. J. Multivariate
  Anal. 113, 59--72.

\bibitem[{Lindsay et~al.(2014)Lindsay, Markatou, and Ray}]{LMR}
Lindsay, B., Markatou, M., Ray, S., 2014. Kernels, degrees of freedom, and
  power of quadratic distance goodness--of--fit tests. J. Amer. Statist. Assoc.
  109, 395--410.

\bibitem[{Mardia et~al.(1979)Mardia, Kent, and Bibby}]{Mardia1979}
Mardia, K.~V., Kent, J.~T., Bibby, J.~M., 1979. Multivariate analysis.
  Probability and Mathematical Statistics: A Series of Monographs and
  Textbooks. Academic Press [Harcourt Brace Jovanovich, Publishers], London-New
  York-Toronto, Ont.

\bibitem[{Meintanis and Hl\'avka(2010)}]{MH10}
Meintanis, S., Hl\'avka, Z., 2010. Goodness-of-fit test for bivariate and
  multivariate skew-normal distributions. Scand. J. Statist. 37, 701--714.

\bibitem[{Meintanis et~al.(2015)Meintanis, Ngatchou-Wandji, and Taufer}]{MNT}
Meintanis, S., Ngatchou-Wandji, J., Taufer, E., 2015. Goodness-of-fit tests for
  multivariate stable distributions based on the empirical characteristic
  function. J. Multivar. Anal. 140, 171--192.

\bibitem[{Meintanis(2005)}]{Meintanis2005}
Meintanis, S.~G., 2005. Permutation tests for homogeneity based on the
  empirical characteristic function. J. Nonparametr. Stat. 17~(5), 583--592.

\bibitem[{Meintanis et~al.(2014)Meintanis, Jim\'{e}nez~Gamero, and
  Alba-fern\'{a}ndez}]{Meintanis2014}
Meintanis, S.~G., Jim\'{e}nez~Gamero, M.~D., Alba-fern\'{a}ndez, V., 2014. A
  class of goodness-of-fit tests based on transformation. Comm. Statist. Theory
  Methods 43~(8), 1708--1735.

\bibitem[{Micchelli et~al.(2006)Micchelli, Xu, and Zhang}]{MXZ}
Micchelli, C., Xu, Y., Zhang, H., 2006. Universal kernels. J. Mach. Learn.
  Resear. 7, 2651--2667.

\bibitem[{Nadarajah(2003)}]{Nadarajah2003}
Nadarajah, S., 2003. The {K}otz-type distribution with applications. Statistics
  37~(4), 341--358.

\bibitem[{Nolan(2013)}]{Nolan2013}
Nolan, J.~P., 2013. Multivariate elliptically contoured stable distributions:
  theory and estimation. Comput. Statist. 28~(5), 2067--2089.

\bibitem[{Sejdinovic et~al.(2013)Sejdinovic, Sriperumbudur, Gretton, and
  Fukumizu}]{SSGF}
Sejdinovic, D., Sriperumbudur, B., Gretton, A., Fukumizu, F., 2013. Equivalence
  of distance--based and {RKHS}--based statistics in hypothesis testing. Ann.
  Statist. 41, 2263--2291.

\bibitem[{{Statisticat} and {LLC.}(2021)}]{LaplacesDemon}
{Statisticat}, {LLC.}, 2021. LaplacesDemon: Complete Environment for Bayesian
  Inference. R package version 16.1.6.

\bibitem[{Sz\'ekely and Rizzo(2005)}]{SR05}
Sz\'ekely, G., Rizzo, M., 2005. A new test for multivariate normality. J.
  Multivar. Anal. 93, 58--80.

\bibitem[{Sz\'ekely and Rizzo(2013)}]{SR13}
Sz\'ekely, G., Rizzo, M., 2013. Energy statistics: A class of statistics based
  on distances. J. Statist. Plann. Infer. 143, 1249--1272.

\bibitem[{Tenreiro(2009)}]{Tenreiro09}
Tenreiro, C., 2009. On the choice of the smoothing parameter for the bhep
  goodness--of--fit test. Comput. Statist. Data Anal. 53, 1038--1053.

\bibitem[{Tenreiro(2019)}]{Tenreiro19}
Tenreiro, C., 2019. On the automatic selection of the tuning parameter
  appearing in certain families of goodness--of--fit tests. J. Statist. Comput.
  Simul. 89, 1780--1797.

\end{thebibliography}


\begin{thebibliography}{4}
\expandafter\ifx\csname natexlab\endcsname\relax\def\natexlab#1{#1}\fi
\expandafter\ifx\csname url\endcsname\relax
  \def\url#1{\texttt{#1}}\fi
\expandafter\ifx\csname urlprefix\endcsname\relax\def\urlprefix{URL }\fi

\bibitem[{Azzalini(2022)}]{sn}
Azzalini, A., 2022. The {R} package \texttt{sn}: The skew-normal and related
  distributions such as the skew-$t$ and the {SUN} (version 2.0.2).
  Universit\`a degli Studi di Padova, Italia, home page:
  \url{http://azzalini.stat.unipd.it/SN/}.

\bibitem[{Fang et~al.(1990)Fang, Kotz, and Ng}]{Fang1990}
Fang, K.~T., Kotz, S., Ng, K.~W., 1990. Symmetric multivariate and related
  distributions. Vol.~36 of Monographs on Statistics and Applied Probability.
  Chapman and Hall, Ltd., London.

\bibitem[{{Statisticat} and {LLC.}(2021)}]{LaplacesDemon}
{Statisticat}, {LLC.}, 2021. LaplacesDemon: Complete Environment for Bayesian
  Inference. R package version 16.1.6.

\bibitem[{Venables and Ripley(2002)}]{MASS}
Venables, W.~N., Ripley, B.~D., 2002. Modern Applied Statistics with S, 4th
  Edition. Springer, New York.

\end{thebibliography}
\bibliographystyle{elsarticle-harv}\biboptions{authoryear}

\newpage
\renewcommand{\baselinestretch}{2}
\renewcommand{\thefootnote}{}
\setcounter{equation}{0}
\setcounter{section}{0}
\setcounter{figure}{0}
\setcounter{page}{0}
\def\theequation{S\arabic{section}.\arabic{equation}}
\def\thesection{S\arabic{section}}
\renewcommand{\thetable}{S.\arabic{table}}
\renewcommand{\thefigure}{S.\arabic{figure}}
\renewcommand{\thepage}{S.\arabic{page}}
$\ $\par \fontsize{12}{14pt plus.8pt minus .6pt}\selectfont

 \centerline{\large\bf Supplementary Material for }
\vspace{2pt} \centerline{\large\bf ``A general Monte Carlo method for multivariate goodness--of--fit }
\vspace{2pt} \centerline{\large\bf testing applied to elliptical families"}
\vspace{.25cm}
 \centerline{Feifei Chen$^{a}$, M. Dolores Jim\'enez--Gamero$^{b}$, Simos Meintanis$^{c,d}$, Lixing Zhu$^{a}$}
\vspace{.4cm}
 \centerline{\it $^a$Center for Statistics and Data Science, Beijing Normal University, Zhuhai, China}
\vspace{2pt}
 \centerline{\it $^b$Department of Statistics and Operations Research, University of Seville, Spain}
\vspace{2pt}
 \centerline{\it $^c$Department of Economics, National and Kapodistrian University of Athens, Athens, Greece}
\vspace{2pt}
 \centerline{\it $^d$Pure and Applied Analytics, North--West University, Potchefstroom, South Africa}

\vspace{.55cm}
\noindent
In this document, random generation for distributions involved and additional simulation results of weight functions different from the main paper are reported.

\section{Generation of random numbers}
In the numerical study section, the generation of random numbers for some multivariate distributions is needed.
We now list the \textsf{R} packages or algorithms involved for readers' convenience.
\begin{itemize}
  \item The multivariate normal distribution can be generated with the function \emph{mvrnorm} from the \textsf{R} package \texttt{MASS} (\citealp{MASS}).

  \item The multivariate Laplace distribution can be generated with the function \emph{rmvl} from the \textsf{R} package \texttt{LaplacesDemon} (\citealp{LaplacesDemon}).

  \item The multivariate Student--t distribution can be generated with the function \emph{rmvt} from the \textsf{R} package \texttt{LaplacesDemon} (\citealp{LaplacesDemon}).

  \item The multivariate skew--t distributions can be generated with the function \emph{rmst} from the \textsf{R} package \texttt{sn} (\citealp{sn}).

  \item We did not find one in the existing \textsf{R} packages that can generate the Kotz--type distribution.
   To simulate random observations from $\mathcal{K}_p(\bb{0}, {\rm{I}}_p, N)$, we use the stochastic representation given by \cite{Fang1990} in Section 3.2.2. That is, $\bb{X}\stackrel{d}{=}R\bb{U}$ when $\bb{X}\sim\mathcal{K}_p(\bb{0}, {\rm{I}}_p, N)$, where $\stackrel{d}{=}$ stands for equality in distribution, $R^2$ follows the gamma distribution $Ga(N+p/2-1, 1/2)$, and $\bb{U}$ is uniformly distributed on the unit sphere and is independent of $R$.
   In particular, the above algorithm can be implemented with the function \emph{my.rmkt} in the \textsf{R} script \emph{Tools.R}, which is also provided as Electric Supplementary Material.
\end{itemize}

\section{Additional simulation studies}

Recall that when we investigating numerically the impact of the resample size $m$, we consider the goodness--of--fit test for multivariate normality and generate data using four different distributions. To be specific, these distributions are $\mathcal{N}_p(\bb{e}_p,\bb{ \Sigma}_{0.5})$, $\mathcal{NM}_p(3)$, $\mathcal{U}^p(0,1)$, and $\mathcal{MAR}_p(Exp)$, Case 1--4 respectively. Please refer to ``Choice of the resample size" in Section 7.1 in the main paper for details.

\begin{figure}[!htb]
 \centering
 \includegraphics [scale=0.52]{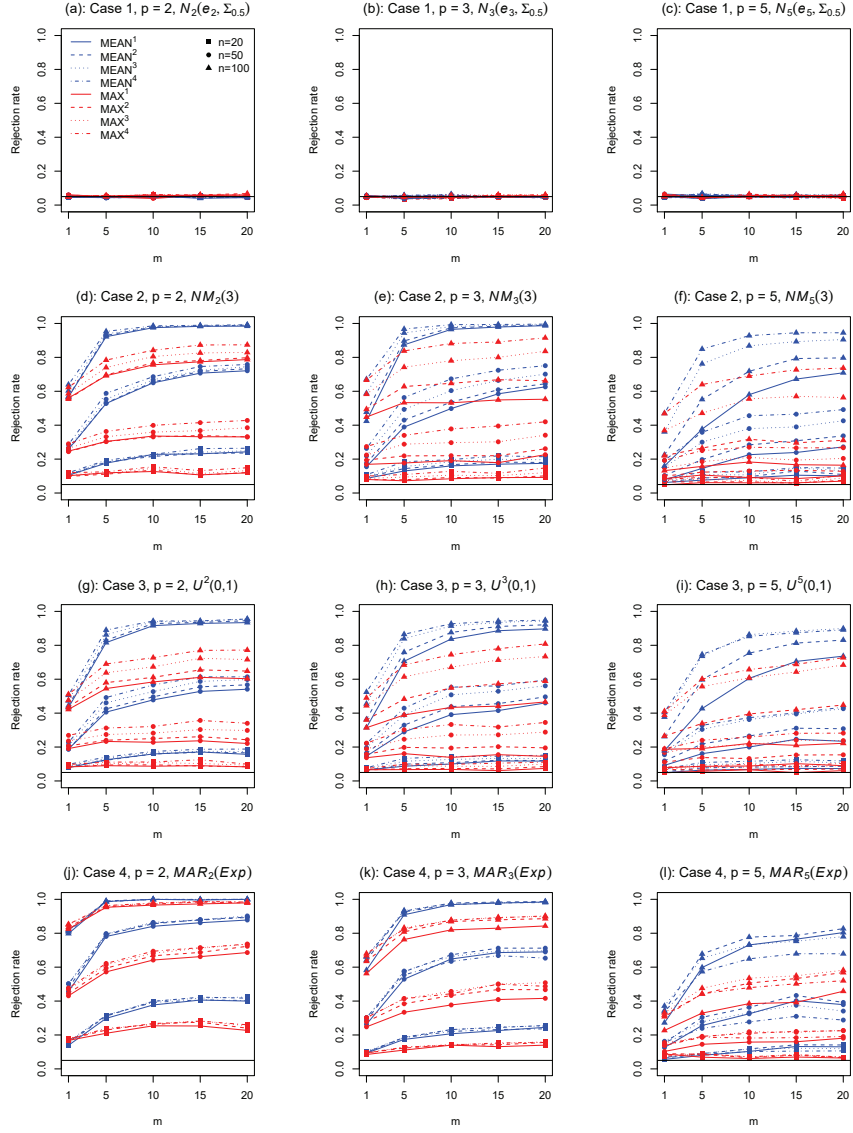}\par
 \vspace{-0.6cm}
 \caption{Empirical rejection rates of $\overline {\widehat T}_{n,m}^{(\Psi)}$ (MEAN) and $\widetilde {\widehat T}_{n,m}^{({\Psi})}$ (MAX) corresponding to the spherical stable weight with $\Psi(\xi)=e^{- \xi^{b/2}}, \ b \in (0,2]$, against different values of $m$ for Cases 1--4. The superscripts 1--4 in the legend correspond to $b=0.5, 1.0, 1.5, 2.0$, respectively. The horizontal line corresponds to the $5\%$ significance level. }  \label{fig-m-Stable}
\end{figure}

\begin{figure}[!htb]
 \centering
 \includegraphics [scale=0.52]{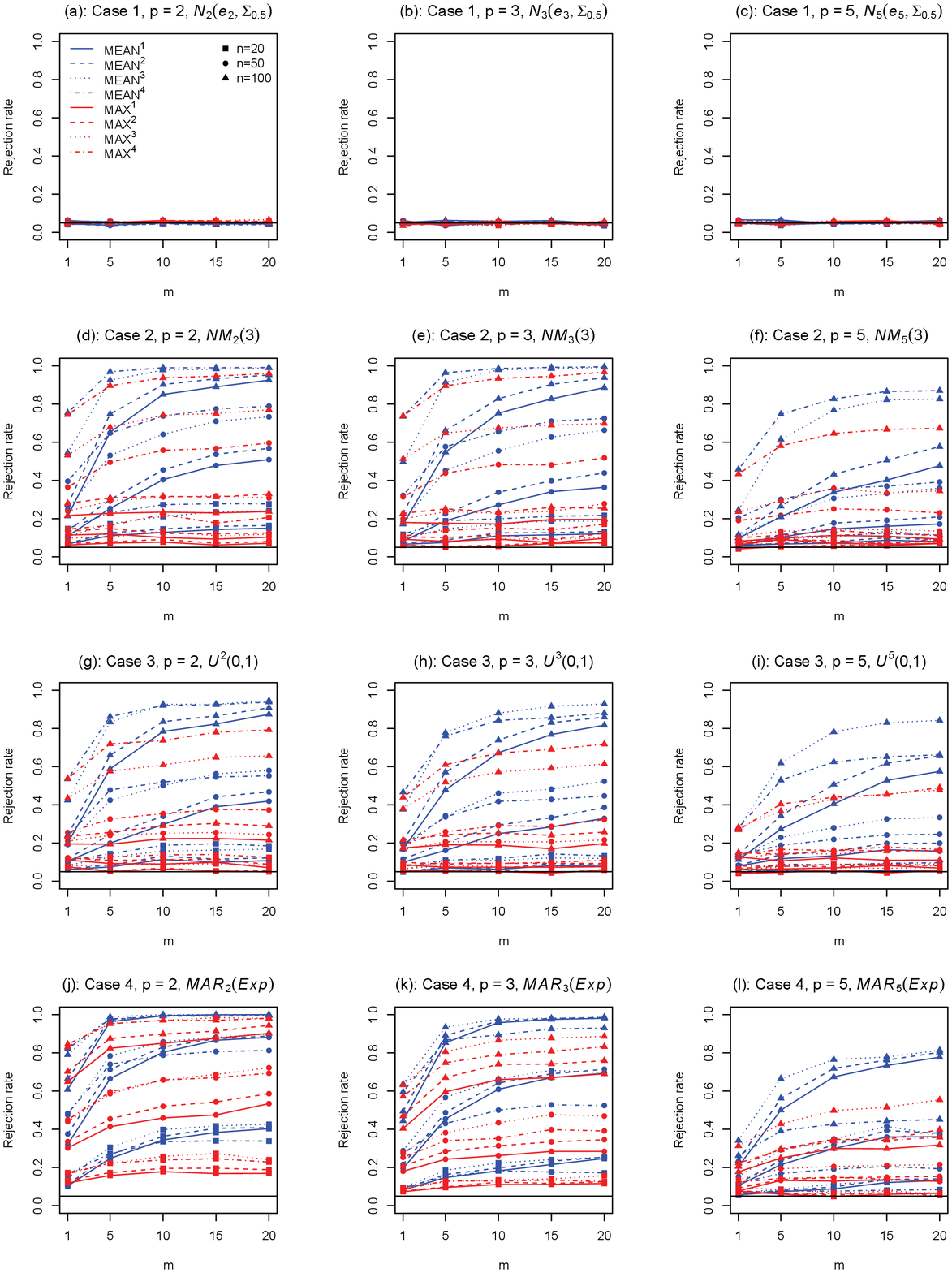}\par
 \vspace{-0.6cm}
 \caption{Empirical rejection rates of $\overline {\widehat T}_{n,m}^{(\Psi)}$ (MEAN) and $\widetilde {\widehat T}_{n,m}^{({\Psi})}$ (MAX) corresponding to the Laplace weight with $\Psi(\xi)=(1+ \xi)^{-b}, \ b>0$, against different values of $m$ for Cases 1--4. The superscripts 1--4 in the legend correspond to $b=0.1, 0.25, 1.0, 4.0$, respectively. The horizontal line corresponds to the $5\%$ significance level. }  \label{fig-m-Laplace}
\end{figure}

Now we consider the densities of spherical stable family with $\Psi(\xi)=e^{- \xi^{b/2}}, \ b \in (0,2]$, and of the generalized Laplace family with $\Psi(\xi)=(1+ \xi)^{-b}, \ b>0$, as weight functions.
For the tuning parameter $b$, we choose $b\in \{0.5, 1.0, 1.5, 2.0\}$ for the stable density and $b\in \{0.1, 0.25, 1.0, 4.0\}$ for the Laplace density.
The empirical rejection rates corresponding to spherical stable weight and Laplace weight against different values of $m$ for Cases 1--4 are displayed in Figures \ref{fig-m-Stable} and \ref{fig-m-Laplace}, respectively.

\end{document}